\documentclass[10pt, oneside, a4paper, abstracton]{scrartcl}

\usepackage{lmodern}
\usepackage[american]{babel}

\usepackage[noadjust]{cite}

\usepackage{amsmath}
\usepackage{amssymb}
\usepackage{amsthm}

\usepackage{tikz}
  \usetikzlibrary{backgrounds}
  \usetikzlibrary{calc}
  \usetikzlibrary{positioning}

\usepackage{pgfplots}
  \pgfplotsset{compat = 1.13,
    colormap name    = viridis,
    unbounded coords = jump
  }
  \usepgfplotslibrary{external}
  \tikzset{external/system call = {%
    lualatex \tikzexternalcheckshellescape
      -halt-on-error
      -interaction=batchmode
      -jobname "\image" "\texsource"}}
  \tikzexternaldisable
  
\usepackage{filemod}

\usepackage{caption}
\usepackage{subcaption}
  
\definecolor{linkBlue}{HTML}{0055C9}
\definecolor{linkRed}{HTML}{FF1A24}
\definecolor{linkPurple}{HTML}{6200D9}

\definecolor{overview}{gray}{0.55}

\definecolor{mpiblue}{HTML}{33A5C3}
\definecolor{mpigreen}{HTML}{007675}
\definecolor{mpired}{HTML}{78004B}
\definecolor{cscorange}{HTML}{FF8F00}

\colorlet{fomcolor}{mpigreen!80}
\colorlet{adtfcolor}{mpiblue!80}
\colorlet{redsubcolor}{cscorange!80}
\colorlet{romcolor}{mpired!80}

\definecolor{plotcolor1}{HTML}{1B9E77}
\definecolor{plotcolor2}{HTML}{D95F02}
\definecolor{plotcolor3}{HTML}{7570B3}
\definecolor{plotcolor4}{HTML}{E7298A}
\definecolor{plotcolor5}{HTML}{66A61E}
\definecolor{plotcolor6}{HTML}{E6AB02}
\definecolor{plotcolor7}{HTML}{A6761D}
\definecolor{plotcolor8}{HTML}{666666}

\usepackage[%
  colorlinks = true,
  linkcolor  = linkBlue,
  citecolor  = linkRed,
  urlcolor   = linkPurple
]{hyperref}
\usepackage{url}
\usepackage{doi}

\usepackage{geometry}
\usepackage{todonotes}

\newcommand{\cP}{\ensuremath{\mathcal{P}}}
\newcommand{\hA}{\ensuremath{\widehat{A}}}
\newcommand{\hB}{\ensuremath{\widehat{B}}}
\newcommand{\hC}{\ensuremath{\widehat{C}}}
\newcommand{\hD}{\ensuremath{\widehat{D}}}
\newcommand{\hE}{\ensuremath{\widehat{E}}}
\newcommand{\hG}{\ensuremath{\widehat{G}}}
\newcommand{\hK}{\ensuremath{\widehat{K}}}
\newcommand{\hM}{\ensuremath{\widehat{M}}}
\newcommand{\hx}{\ensuremath{\hat{x}}}
\newcommand{\hy}{\ensuremath{\hat{y}}}
\newcommand{\tA}{\ensuremath{\widetilde{A}}}
\newcommand{\tE}{\ensuremath{\widetilde{E}}}
\newcommand{\tT}{\ensuremath{\widetilde{T}}}
\newcommand{\tW}{\ensuremath{\widetilde{W}}}

\newcommand{\trans}{\ensuremath{\mkern-1.5mu\mathsf{T}}}

\DeclareMathOperator{\disk}{disk}
\DeclareMathOperator{\sign}{sign}
\DeclareMathOperator{\trace}{tr}

\def\mvdots{\vbox{\baselineskip=4pt \lineskiplimit=0pt 
  \kern3pt \hbox{.}\hbox{.}\hbox{.}}}
\def\mddots{\mathinner{\mkern1mu\raise7pt\vbox{\kern3.5pt\hbox{.}}\mkern2mu 
  \raise4pt\hbox{.}\mkern2mu\raise1pt\hbox{.}\mkern1mu}}

\newcommand{\morlab}{\mbox{MORLAB}}

\pgfplotscreateplotcyclelist{plotlist}{
  {black, solid, line width = 1pt},
  {plotcolor1, mark options = {solid}, mark = *, mark size = 1.5pt,
    mark repeat = 80, mark phase = 0, line width = 1pt},
  {plotcolor2, mark options = {solid}, mark = o, mark size = 1.5pt,
    mark repeat = 80, mark phase = 8, line width = 1pt},
  {plotcolor3, mark options = {solid}, mark = x, mark size = 1.5pt,
    mark repeat = 80, mark phase = 16, line width = 1pt},
  {plotcolor4, mark options = {solid}, mark = +, mark size = 1.5pt,
    mark repeat = 80, mark phase = 24, line width = 1pt},
  {plotcolor5, mark options = {solid}, mark = triangle*, mark size = 1.5pt,
    mark repeat = 80, mark phase = 32, line width = 1pt},
  {plotcolor6, mark options = {solid}, mark = square*, mark size = 1.5pt,
    mark repeat = 80, mark phase = 40, line width = 1pt},
  {plotcolor7, mark options = {solid}, mark = diamond*, mark size = 1.5pt,
    mark repeat = 80, mark phase = 48, line width = 1pt},
  {plotcolor8, mark options = {solid}, mark = Mercedes star, mark size = 1.5pt,
    mark repeat = 80, mark phase = 56, line width = 1pt}
}

\def\formtmp#1#2{{\vskip12pt\noindent\fboxsep=0pt\colorbox{#1}{%
\vbox{\vskip3pt\hbox to \textwidth{\hskip3pt\vbox{%
\raggedright\noindent\textbf{#2\vphantom{Qy}}}\hfill}\vspace*{3pt}}}%
\par\vskip2pt\noindent\kern0pt}}

\usepackage{tabularx}
\usepackage{ltablex}

\newenvironment{overview}[1]{\ignorespaces\def\stmtopen##1{##1}%
\formtmp{overview}{#1}}{\par\noindent\textcolor{overview}{%
\rule{\columnwidth}{1pt}}\vskip2pt\par\addvspace{\baselineskip}}%

\theoremstyle{definition}\newtheorem{remark}{Remark}

\begin{document}


\title{MORLAB -- The Model Order Reduction LABoratory}

\author{%
  Peter~Benner\thanks{
    Max Planck Institute for Dynamics of Complex Technical Systems,
    Sandtorstr. 1, 39106 Magdeburg, Germany.\newline
    E-mail: \texttt{\href{mailto:benner@mpi-magdeburg.mpg.de}%
      {benner@mpi-magdeburg.mpg.de}}
    \newline
    Otto von Guericke University, Faculty of Mathematics,
    Universit{\"a}tsplatz 2, 39106 Magdeburg, Germany.\newline
    E-mail: \texttt{\href{mailto:peter.benner@ovgu.de}%
      {peter.benner@ovgu.de}}} \and
  Steffen~W.~R.~Werner\thanks{
    Max Planck Institute for Dynamics of Complex
    Technical Systems, Sandtorstr. 1, 39106 Magdeburg, Germany.\newline
    E-mail: \texttt{\href{mailto:werner@mpi-magdeburg.mpg.de}%
      {werner@mpi-magdeburg.mpg.de}}}
}

\date{~}
    
\maketitle

\begin{abstract}
  For an easy use of model order reduction techniques in applications, software
  solutions are needed.
  In this paper, we describe the \morlab{}, Model Order Reduction LABoratory,
  toolbox as an efficient implementation of model reduction techniques for
  dense, medium-scale linear time-invariant systems.
  Giving an introduction to the underlying programming principles of the
  toolbox, we show the basic idea of spectral splitting and present an overview
  about implemented model reduction techniques.
  Two numerical examples are used to illustrate different use cases of the
  \morlab{} toolbox.
\end{abstract}


\section{Introduction}%
\label{sec:intro}

For the modeling of natural processes as, e.g., fluid dynamics, chemical
reactions or the behavior of electronic circuits, power or gas transportation
networks, dynamical input-output systems are used
\begin{align} \label{eqn:sys}
  G: \left\{
  \begin{aligned}
    0 & = f(x(t), Dx(t), \ldots, D^{k}x(t), u(t)),\\
    y(t) & = h(x(t), Dx(t), \ldots, D^{k}x(t), u(t)),
  \end{aligned}\right.
\end{align}
with states $x(t) \in \mathbb{R}^{n}$, inputs $u(t) \in \mathbb{R}^{p}$ and
outputs $y(t) \in \mathbb{R}^{p}$.
The operator $D^{j}$ denotes the derivative or shift operator of order
$j \in \mathbb{N}$ in case of underlying continuous- or discrete-time dynamics.
Due to the demand for increasing the accuracy of models, the number of states
describing~\eqref{eqn:sys} is drastically increasing and, consequently,
there is a high demand for computational resources (time and memory) when
using~\eqref{eqn:sys} in simulations or controller design.
A solution to this problem is given by model order reduction, which aims for the
construction of a surrogate model $\hG$, with a much smaller number of internal
states $\hx(t) \in \mathbb{R}^{r}$, $r \ll n$, which approximates the
input-to-output behavior of~\eqref{eqn:sys} such that
\begin{align*}
  \lVert y - \hy \rVert & \leq \mathrm{tol} \cdot \rVert u \lVert,
\end{align*}
for an appropriately defined norm, a given tolerance $\mathrm{tol}$ and all
admissible inputs $u$, where $\hy$ is the output of the reduced-order system.

\begin{table}[!t]
  \caption{Code meta data of the latest \morlab{} version~\cite{morBenW19b}.}
  \label{tab:meta_data}
  \centering
  \begin{tabular}{rl}
    \hline\noalign{\smallskip}
    name (shortname) & Model Order Reduction LABoratory (\morlab{})\\
    version (release-date) & 5.0 (2019-08-23)\\
    identifier (type) & \href{https://doi.org/10.5281/zenodo.3332716}%
      {doi: 10.5281/zenodo.3332716} (doi)\\
    authors & Peter Benner, Steffen W. R. Werner\\
    orcids & \href{https://orcid.org/0000-0003-3362-4103}{0000-0003-3362-4103},
      \href{https://orcid.org/0000-0003-1667-4862}{0000-0003-1667-4862}\\
    topic (type) & Model Reduction (Toolbox)\\
    license (type) & GNU Affero General Public License v3.0 (open)\\
    languages & MATLAB\\
    dependencies & MATLAB ($\geq$ 2012b), Octave ($\geq$ 4.0.0)\\
    systems & Linux, MacOS, Windows\\
    website & \url{http://www.mpi-magdeburg.mpg.de/projects/morlab}\\
    \noalign{\smallskip}\hline\noalign{\smallskip}
  \end{tabular}
\end{table}

A software solution for model order reduction of dynamical systems is the
\textbf{\morlab{}}, \textbf{M}odel \textbf{O}rder \textbf{R}eduction
\textbf{LAB}oratory, toolbox.
Originating from~\cite{morBen06a}, the toolbox is mainly developed as efficient
open source implementation of established matrix equation-based model reduction
methods.
Nowadays, it is one of the most efficient model reduction toolboxes for 
dense, medium-scale, linear time-invariant systems, with an implementation
compatible with MathWorks MATLAB and GNU Octave.
In the latest version~\cite{morBenW19b}, \morlab{} gives a large variety of
balancing-based model reduction methods and also some non-projective methods.
Most of those are not known to be implemented somewhere else.
In contrast to other software solutions, the general philosophy of \morlab{}
is to work on invariant subspaces rather than with spectral decompositions or
projections on hidden manifolds, which results in fast and accurate
implementations.
Mainly the two spectral projection methods, the matrix sign function and the
right matrix pencil disk function, are used in the underlying implementations.
Therefore, \morlab{} is suited as backend source code for multi-step model
reduction approaches, for example, using a pre-reduction step;
see, e.g.,~\cite{morLehE07, morSaaSW19}.
Additionally to model order reduction methods, the toolbox implements efficient
matrix equation solvers, system-theoretic subroutines and evaluation routines to
examine original and reduced-order systems in the frequency and time
domain.
Due to the brevity of the paper, the additional main features are not 
further considered in detail.

In this paper, we will describe the underlying principles and structures of the
\morlab{} toolbox and give some applications of the software.
The meta data of the latest \morlab{} version~\cite{morBenW19b} can be found in
Table~\ref{tab:meta_data}.
In the following, Section~\ref{sec:design} starts with an introduction of the
programming principles that were used in \morlab{}.
Afterwards, Section~\ref{sec:specdecomp} gives the underlying ideas of
the spectral splitting, on which the toolbox bases, followed by
Section~\ref{sec:mormeths} with an overview about the implemented model
reduction methods.
In Section~\ref{sec:examples}, two applications of using \morlab{} as backend
software are presented.
The paper is concluded by Section~\ref{sec:conclusions}.


\section{Code design principles}%
\label{sec:design}

\begin{figure}[!t]
  \begin{center}
    \resizebox{\textwidth}{!}{
  \tikzexternalenable%
  \tikzsetnextfilename{morlab_workflow}%
  \filemodCmp{graphics/morlab_workflow.tikz}{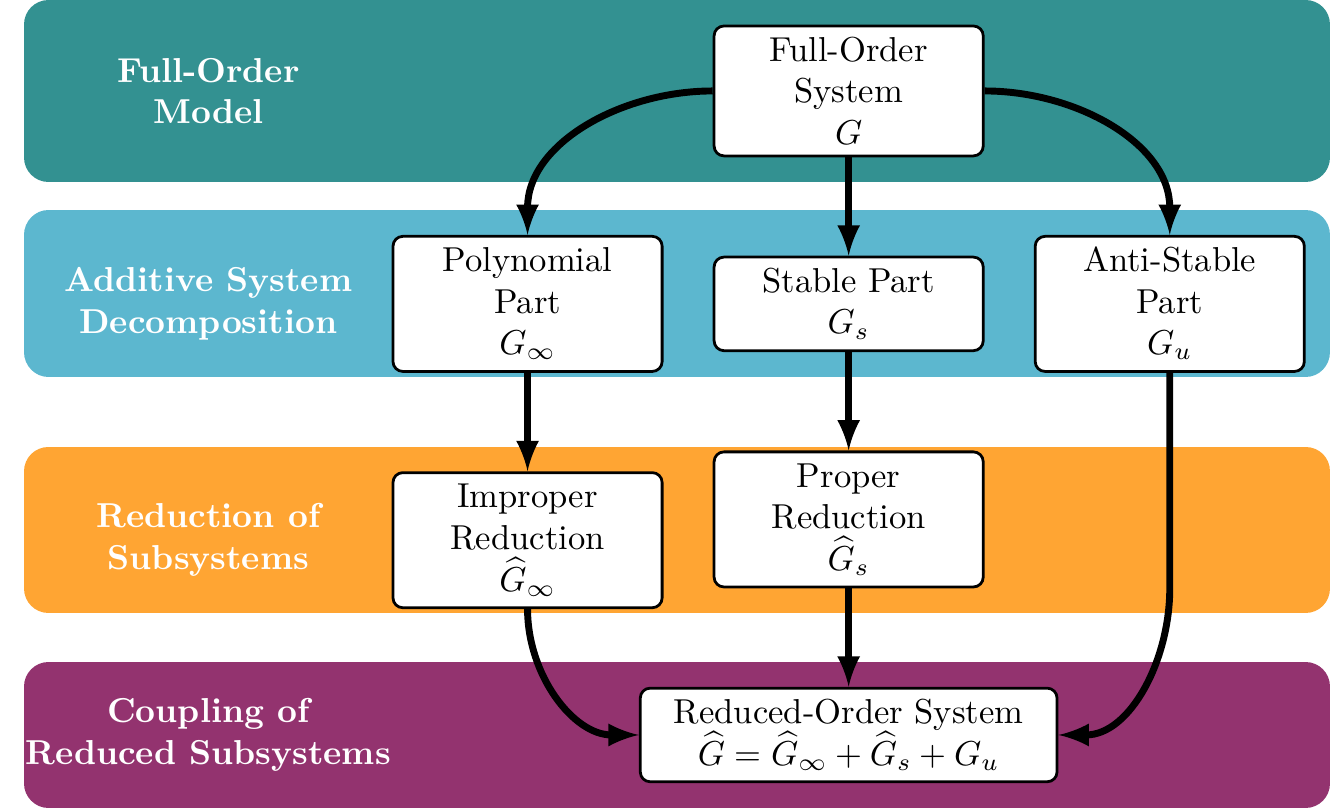}%
    {\tikzset{external/remake next}}{}%
%
%
%

\begin{tikzpicture}
  \tikzstyle{snode} = [
    draw = black,
    fill = white,
    thick,
    rounded corners = 0.1cm
  ]
  
  \tikzstyle{arr} = [
    -latex,
    line width = 2pt,
    draw       = black
  ]
  
  \node[snode](G){\parbox{2.5cm}
    {\centering Full-Order System\\$G$}};
      
  \node[snode](Gsp)[below = 1cm of G.south]
    {\parbox{2.5cm}{\centering Stable Part\\$G_{s}$}};
  \node[snode](P)[left = .5cm of Gsp.west]
    {\parbox{2.5cm}{\centering Polynomial Part\\$G_{\infty}\vphantom{G_{s}}$}}; 
  \node[snode](Gu)[right = .5cm of Gsp.east]
    {\parbox{2.5cm}{\centering Anti-Stable Part\\$G_{u}$}};
      
  \node[snode](Gspr)[below = 1cm of Gsp.south]
    {\parbox{2.5cm}{\centering Proper Reduction\\$\hG_{s}$}};
  \node[snode](Pr)[below = 1cm of P.south]
    {\parbox{2.5cm}{\centering Improper Reduction\\$\hG_{\infty}
    \vphantom{\hG_{s}}$}};
      
  \node[snode](Gr)[below = 1cm of Gspr.south]
    {\parbox{4cm}{\centering Reduced-Order System\\
    $\hG = \hG_{\infty} + \hG_{s} + G_{u}$}};
      
  \draw[arr] (G.west) to[in = 90, out = 180] (P.north);
  \draw[arr] (G.south) -- (Gsp.north);
  \draw[arr] (G.east) to[in = 90, out = 0] (Gu.north);
    
  \draw[arr] (P.south) -- (Pr.north);
  \draw[arr] (Gsp.south) -- (Gspr.north);
    
  \draw[arr]
    (Gu.south) -- (Gspr.south -| Gu.south) to[in = 0, out = 270] (Gr.east);
  \draw[arr] (Pr.south) to[in = 180, out = 270] (Gr.west);
  \draw[arr] (Gspr.south) -- (Gr.north);
      
  \node(phase1)[left = 3.5cm of G.west, text = white]
    {\parbox{3cm}{\centering \textbf{Full-Order Model}}};
      
  \node(phase2)[text = white] at (phase1.south |- P.west)
    {\parbox{4cm}{\centering \textbf{Additive System\\
    Decomposition}}};
      
  \node(phase3)[text = white] at (phase1.south |- Pr.west)
    {\parbox{4cm}{\centering \textbf{Reduction of\\Subsystems}}};
      
  \node(phase4)[text = white] at (phase1.south |- Gr.west)
    {\parbox{4cm}{\centering \textbf{Coupling of\\Reduced Subsystems}}};
      
  \begin{pgfonlayer}{background}
    \filldraw [line width = .5cm, join = round, fomcolor]
      (phase1.west |- G.north) rectangle (Gu.east |- G.south);
        
    \filldraw [line width = .5cm, join = round, adtfcolor]
      (phase1.west |- P.north) rectangle (Gu.east |- Gsp.south);
        
    \filldraw [line width = .5cm, join = round, redsubcolor]
      (phase1.west |- Pr.north) rectangle (Gu.east |- Gspr.south);
        
    \filldraw [line width = .5cm, join = round, romcolor]
      (phase1.west |- Gr.north) rectangle (Gu.east |- Gr.south);
  \end{pgfonlayer}
\end{tikzpicture}%
  \tikzexternaldisable%
}
    \caption{General \morlab{} workflow.}
    \label{fig:morlab_workflow}
  \end{center}
\end{figure}

The main aim of the \morlab{} toolbox is to give efficient and comparable
implementations of many different model reduction methods.
Following certain design principles, which will be explained in more detail
in the upcoming subsections, the following list of main features briefly
summarizes the \morlab{} toolbox.

\begin{overview}{Feature checklist}
  {\def\arraystretch{1.3}
  \begin{tabularx}{.96\textwidth}{lX}
    Open source and free & 
      The toolbox is running under the GNU Affero
      General Public License v3.0 and is freely available on the project
      website and on Zenodo. \\
    Fast and exact & Using spectral projection methods, the toolbox can
      outperform other established software in terms of accuracy and speed. \\
    Unified framework & All model reduction routines share the same
      interface and allow for quick exchange and easy comparison between the
      methods. \\
    Modular & Each subroutine can be called on its own by the user to be
      used and combined in varies ways. \\
    Portable & No binary extensions are required, which allows for running
      the toolbox with bare MATLAB or Octave installations.
  \end{tabularx}}
  \smallskip
  
\end{overview}

In general, \morlab{} uses spectral projection methods for all steps of the
model reduction procedure.
Fig.~\ref{fig:morlab_workflow} shows the different stages in \morlab{}
from the full-order to the reduced-order model.
First, the full-order model is decomposed into (at most) three subsystems that
can usually be considered independently of each other for the application of
model reduction techniques.
This first main step, the additive system decomposition, is discussed in more
detail in Section~\ref{sec:specdecomp}.
Afterwards, the model reduction methods are applied to the resulting subsystems.
An overview of those can be found in Section~\ref{sec:mormeths}.
At the end, the reduced subsystems are coupled for the resulting reduced-order
model.
Based on this basic workflow, the different design principles applied in
\morlab{} are explained in the following.
For the sake of brevity, mainly the model reduction routines are considered.


\subsection{Toolbox structure}

The routines in \morlab{} follow a strict structure and naming scheme to make
them easy to find and interpret in terms of their objective.
Describing first the general structure, the routines of the toolbox are divided
by their purpose into the following subdirectories:
{\def\arraystretch{1.3}
\begin{tabularx}{\textwidth}{lX}
  \texttt{\textbf{checks/}} &
    Contains subroutines that are used for internal checks of data, e.g.,
    if the system structures fit to the model reduction methods. \\
  \texttt{\texttt{\textbf{demos/}}} &
    Contains example scripts showing step-by-step explanations of the different
    main features of the toolbox. \\
  \texttt{\textbf{eqn\_solvers/}} &
    Contains the matrix equation solvers. \\
  \texttt{\textbf{evaluation/}} &
    Contains functions to evaluate the full-order or reduced-order models in the
    time or frequency domain. \\
  \texttt{\textbf{mor/}} &
    Contains the model reduction routines. \\
  \texttt{\textbf{subroutines/}} &
    Contains auxiliary and system-theoretic routines that are used by the model
    reduction techniques, matrix equation solvers or evaluation functions.
\end{tabularx}}
\smallskip

{\setlength\tabcolsep{1.5pt}
\begin{table}[!t]
  \caption{Currently supported system classes.}
  \label{tab:sys}
  \centering
  \begin{tabular}{lcc}
    \multicolumn{1}{c}{\textbf{Class}} & 
      \multicolumn{1}{c}{\textbf{System equations}} &
      \multicolumn{1}{c}{\textbf{Routine name}}\\
    \hline\noalign{\medskip}
    Continuous-time standard systems & 
      \begin{minipage}{.26\textwidth}
      \vspace{-\baselineskip}
      {\begin{align*}
        \begin{aligned}
          \dot{x}(t) & = Ax(t) + Bu(t),\\[-.25\baselineskip]
           y(t) & = Cx(t) + Du(t)
        \end{aligned}
      \end{align*}}
      \vspace{-\baselineskip}
      \end{minipage} &
      \texttt{ct\_ss}\\ \noalign{\medskip}
    Discrete-time standard systems &
      \begin{minipage}{.27\textwidth}
      \vspace{-\baselineskip}
      {\begin{align*}
        \begin{aligned}
          x_{k+1} & = Ax_{k} + Bu_{k},\\[-.25\baselineskip]
          y_{k} & = Cx_{k} + Du_{k}
        \end{aligned}
      \end{align*}}
      \vspace{-\baselineskip}
      \end{minipage} &
      \texttt{dt\_ss}\\ \noalign{\medskip}
    Continuous-time descriptor systems & 
      \begin{minipage}{.27\textwidth}
      \vspace{-\baselineskip}
      {\begin{align*}
        \begin{aligned}
          E\dot{x}(t) & = Ax(t) + Bu(t),\\[-.25\baselineskip]
           y(t) & = Cx(t) + Du(t)
        \end{aligned}
      \end{align*}}
      \vspace{-\baselineskip}
      \end{minipage} &
      \texttt{ct\_dss}\\ \noalign{\medskip}
    Discrete-time descriptor systems &
      \begin{minipage}{.28\textwidth}
      \vspace{-\baselineskip}
      {\begin{align*}
        \begin{aligned}
          Ex_{k+1} & = Ax_{k} + Bu_{k},\\[-.25\baselineskip]
          y_{k} & = Cx_{k} + Du_{k}
        \end{aligned}
      \end{align*}}
      \vspace{-\baselineskip}
      \end{minipage} &
      \texttt{dt\_dss}\\ \noalign{\medskip}
    Continuous-time second-order systems & 
      \begin{minipage}{.41\textwidth}
      \vspace{-\baselineskip}
      {\begin{align*}
        \begin{aligned}
          M\ddot{x}(t) & = -Kx(t) - E\dot{x}(t)
            + B_{u}u(t),\\[-.25\baselineskip]
          y(t) & = C_{p}x(t) + C_{v}\dot{x}(t) + Du(t)
        \end{aligned}
      \end{align*}}
      \vspace{-\baselineskip}
      \end{minipage} &
      \texttt{ct\_soss}\\
    \noalign{\medskip}\hline\noalign{\smallskip}
  \end{tabular}
\end{table}}

Considering to the naming scheme of \morlab{}, each function starts with
\texttt{ml\_} as assignment to the toolbox.
This makes \morlab{} routines easier to distinguish from other source codes and
also allows for easy searching.
Mainly the model reduction routines, but also some subroutines are additionally
named after the system classes they can be applied to.
Currently, there are routines for continuous- (\texttt{ct}) and discrete-time
(\texttt{dt}) dynamical system with equations that describe standard
(\texttt{ss}), descriptor (\texttt{dss}) or second-order state spaces
(\texttt{soss}).
The resulting different system classes, supported in the latest \morlab{}
version, are summarized in Table~\ref{tab:sys} with their names, system
equations and the corresponding naming schemes.


\subsection{Function interfaces}

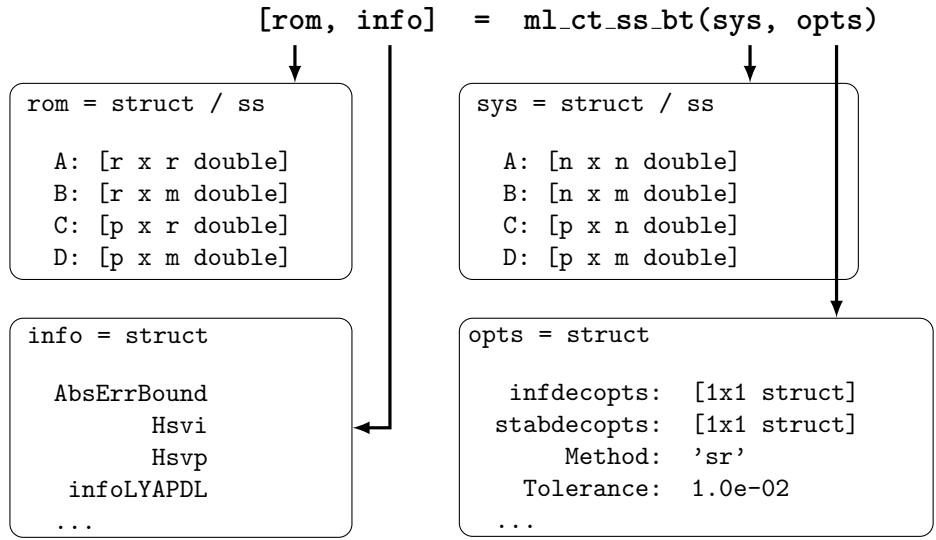
\begin{figure}[!t]
  \begin{center}
  \tikzexternalenable%
  \tikzsetnextfilename{morlab_func}%
  \filemodCmp{graphics/morlab_func.tikz}{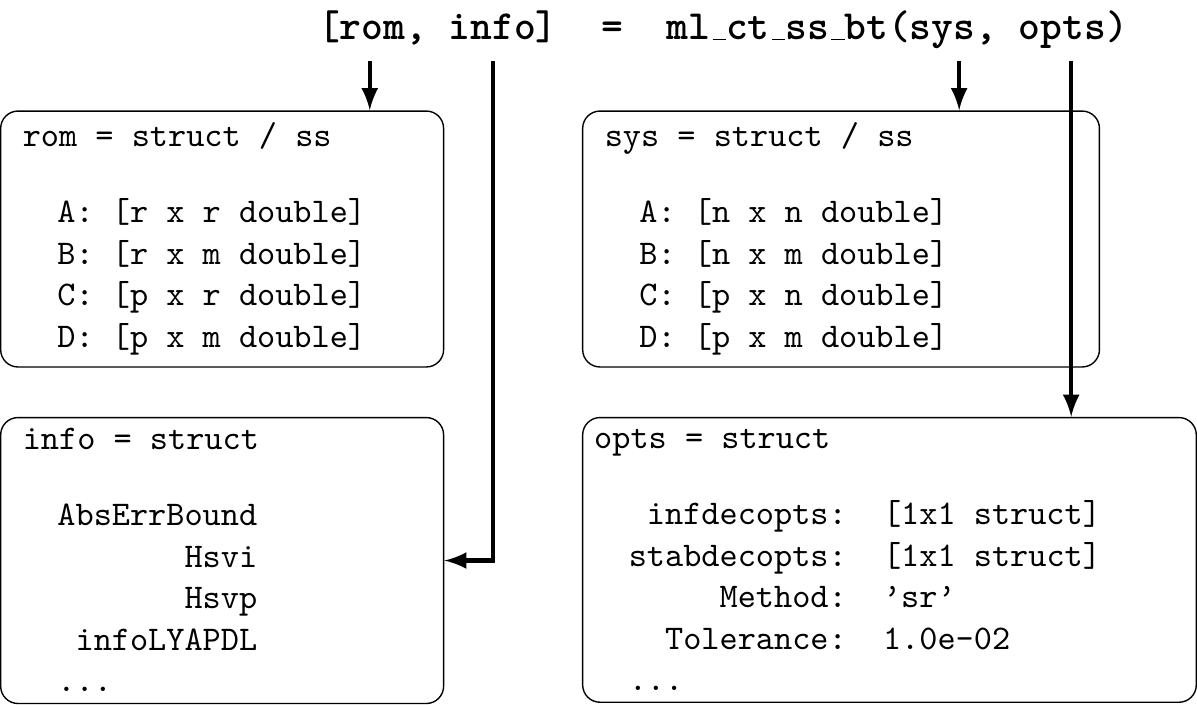}%
    {\tikzset{external/remake next}}{}%
  \begin{tikzpicture}
  \node(bt){\large\textbf{\texttt{[rom, info]~~=~~ml\_ct\_ss\_bt(sys, opts)}}};
      
  \node[
    draw            = black,
    minimum width   = .35\textwidth,
    minimum height  = 2.6cm,
    rounded corners = .5em
  ](sys)[below = .5cm of bt.south, anchor = north, xshift = 1.2cm]{
    \begin{minipage}{.32\textwidth}
      \texttt{sys = struct / ss}\\[1em]
      \hspace*{1em}\texttt{A: [n x n double]}\\
      \hspace*{1em}\texttt{B: [n x m double]}\\
      \hspace*{1em}\texttt{C: [p x n double]}\\
      \hspace*{1em}\texttt{D: [p x m double]}
     \end{minipage}};
        
  \node[
    draw            = black,
    minimum width   = .35\textwidth,
    minimum height  = 2.6cm,
    rounded corners = .5em
  ](opts)[below = .5cm of sys.south west, anchor = north west]{
    \begin{minipage}{.4\textwidth}
      \texttt{opts = struct}\\[1em]
      \hspace*{1em}\texttt{\phantom{a}infdecopts: [1x1 struct]}\\
      \hspace*{1em}\texttt{stabdecopts: [1x1 struct]}\\
      \hspace*{1em}\texttt{\phantom{aaaaa}Method: 'sr'}\\
      \hspace*{1em}\texttt{\phantom{aa}Tolerance: 1.0e-02}\\
      \hspace*{1em}\texttt{...}
    \end{minipage}};
        
  \node[
    draw            = black,
    minimum width   = .3\textwidth,
    minimum height  = 2.6cm,
    rounded corners = .5em
  ](rom)[below = .5cm of bt.south west, anchor = north, xshift = -.85cm]{
    \begin{minipage}{.27\textwidth}
      \texttt{rom = struct / ss}\\[1em]
      \hspace*{1em}\texttt{A: [r x r double]}\\
      \hspace*{1em}\texttt{B: [r x m double]}\\
      \hspace*{1em}\texttt{C: [p x r double]}\\
      \hspace*{1em}\texttt{D: [p x m double]}
    \end{minipage}};
        
  \node[
    draw            = black,
    minimum width   = .3\textwidth,
    minimum height  = 2.6cm,
    rounded corners = .5em
  ](info)[below = .5cm of rom.south, anchor = north]{
    \begin{minipage}{.27\textwidth}
      \texttt{info = struct}\\[1em]
      \hspace*{1em}\texttt{AbsErrBound}\\
      \hspace*{1em}\texttt{\phantom{aaaaaaa}Hsvi}\\
      \hspace*{1em}\texttt{\phantom{aaaaaaa}Hsvp}\\
      \hspace*{1em}\texttt{\phantom{a}infoLYAPDL}\\
      \hspace*{1em}\texttt{...}
    \end{minipage}};

  \draw[-latex, very thick] (bt.south) ++(2.4,0) 
    -- ++(0,-.5);
  \coordinate(tmp) at ($(bt.south east)+(-.7,0)$); 
  \draw[-latex, very thick] (bt.south east) ++(-.7,0)
    -- (tmp |- opts.north);
  \draw[-latex, very thick] (bt.south west) ++(1.9,0) |- (info.east);
  \draw[-latex, very thick] (bt.south west) ++(.65,0)
    -- ($(rom.north) + (1.5,0)$);
\end{tikzpicture}%
  \tikzexternaldisable%

    \caption{Example function call of a model reduction routine in \morlab{}.}
    \label{fig:morlab_func}
  \end{center}
\end{figure}

A typical function call in \morlab{} can be seen in Fig.~\ref{fig:morlab_func}.
From before, we know that the called function is a \morlab{} routine for
continuous-time standard systems (see Table~\ref{tab:sys}).
The actual function name, \texttt{bt}, stands for the balanced truncation
method.
Fig.~\ref{fig:morlab_func} shows the principle idea in \morlab{} to give an
easy interface to the user.
Here, \texttt{sys} contains the data of the original system, while \texttt{rom}
gives the resulting reduced-order model in exactly the same format as the
original model was given, indicating the purpose of using reduced-order
models as surrogates for the original system.
In general, \morlab{} supports three different interfaces for model reduction
methods.
It is possible to pass directly the system matrices to the function
(e.g., \texttt{ml\_ct\_ss\_bt(A, B, C, D, opts)}) or to construct the system as
an object by using the native data type \texttt{struct}, with appropriate naming
of fields, or the state-space object (\texttt{ss}) introduced by the Control
System Toolbox\texttrademark{} in MATLAB or the 'control' package in Octave.
The latter format allows for easy interconnection to other model reduction
software and also for using system-theoretic routines implemented in the two
mentioned software libraries.

The second important part of the \morlab{} interface for nearly all routines
are the \texttt{opts} and \texttt{info} structs, as shown in
Fig.~\ref{fig:morlab_func}.
Supporting the feature of configurability, the \texttt{opts} struct allows
the user the rearrangement of all computational parameters, which would
be usually set by the function itself during runtime.
In general, each \morlab{} function that allows the user to change optional
parameters for the computations has an \texttt{opts} struct for that purpose.
As result, higher level routines can contain nested structs to change
computational parameters of used subroutines.
Fig.~\ref{fig:morlab_opts} shows an example \texttt{opts} struct for the 
\texttt{ml\_ct\_ss\_bt}.
This struct again contains entries ending on \texttt{opts} denoting also
\texttt{opts} structs for subroutines that are called by the main function.
Beside changing computational parameters, a second aim of the \texttt{opts}
struct is the a priori determination of system information.
For example, if a system is known to be stable, the additive decomposition into
the stable and anti-stable subsystems can be turned off using the
\texttt{opts} struct to avoid unnecessary computations.
For easy application, only entries, which the user wants to change, need to be
existing in the struct.
Also, the toolbox comes with an option constructor (\texttt{ml\_morlabopts}),
which creates a complete but empty \texttt{opts} struct for a given function
name.
The consistent naming of optional parameters between different routines allows
the easy reuse of \texttt{opts} structs for different functions.

The counterpart of the \texttt{opts} struct is the \texttt{info} struct.
Here, information about the performance and results of the routine are 
collected.
As for \texttt{opts}, the \texttt{info} struct can be nested as it contains
structs starting with \texttt{info}, which give information about used
subroutines.
Also, this struct is used for optional outputs, e.g., projection matrices of
a model reduction method can be stored in here.


\subsection{Documentation}

\morlab{} comes with an extensive documentation that is accessible in several
ways.
Each routine has a complete inline documentation, which can be displayed by the
\texttt{help} command, containing the syntax, description and literature
references for background information.
Besides, a complete overview about the existing \morlab{} routines with short
description can be generated by \texttt{help morlab}.
As usual for MATLAB toolboxes, a full HTML documentation is provided in the 
toolbox and demo scripts can be used as a starting how-to to get into the main
features of the toolbox.

\begin{figure}[!t]
  \begin{center}
    \input{graphics/morlab_opts.vbt}
  \end{center}
  
  \caption{Example \texttt{opts} struct for the \texttt{ml\_ct\_ss\_bt}
    function.}
  \label{fig:morlab_opts}
\end{figure}


\section{Additive system decomposition approach}%
\label{sec:specdecomp}

Most model order reduction methods are in a certain sense restricted with
respect to the spectrum of the underlying system matrices, e.g., the classical
balanced truncation method can only be applied to first-order systems with
finite stable matrix pencils.
Other software solutions use therefor either an eigendecomposition of the
system matrices in the beginning or apply projections onto the hidden manifolds.
In \morlab{}, this problem is solved by working directly with the corresponding
invariant subspaces of the matrix pencil.
As shown in Fig.~\ref{fig:morlab_workflow}, this results in the additive
decomposition of the full-order system into independent reducable subsystems,
in the literature known as additive decomposition of the transfer function,
which will be coupled at the end again.
\morlab{} has two different approaches for this additive decomposition based on
either the solution of a Sylvester equation or on a block wise projection
approach.
This gives \morlab{} the advantage of handling unstructured systems, while
staying efficient and accurate due to only computing the necessary deflating
subspaces.
For both approaches, the matrix sign and disk functions are used, as quickly
defined below.

Let $Y \in \mathbb{R}^{n \times n}$ be a matrix with no purely imaginary 
eigenvalues, then the Jordan canonical form of $Y$ can be written as
\begin{align} \label{eqn:jordanform}
  Y & = S \begin{bmatrix} J_{-} & 0 \\ 0 & J_{+} \end{bmatrix} S^{-1},
\end{align}
where $S$ is an invertible transformation matrix, $J_{-}$ contains the $k$ 
eigenvalues of $Y$ with negative real parts and $J_{+}$ the $n - k$
eigenvalues with positive real parts.
The \emph{matrix sign function} is then defined as
\begin{align} \label{eqn:sign}
  \sign(Y)  & = S \begin{bmatrix} -I_{k} & 0 \\ 0 & I_{n - k}
    \end{bmatrix} S^{-1},
\end{align}
with $S$ the transformation matrix from~\eqref{eqn:jordanform}; see,
e.g.,~\cite{morRob80}.
Efficient computations can be based on a Newton scheme.

Let $\lambda X - Y$, with $X, Y \in \mathbb{R}^{n \times n}$, be a regular
matrix pencil with no eigenvalues on the unit circle and its Weierstrass
canonical form be written as
\begin{align} \label{eqn:weierform}
  \lambda X - Y & = W \begin{bmatrix} \lambda I_{k} - J_{0} & 0 \\
    0 & \lambda N - J_{\infty} \end{bmatrix} T^{-1},
\end{align}
where $W, T$ are invertible transformation matrices, $\lambda I_{k} - J_{0}$ 
contains the $k$ eigenvalues inside the unit disk and $\lambda N - J_{\infty}$
the $n - k$ eigenvalues outside the unit disk.
The \emph{right matrix pencil disk function} is then defined by
\begin{align} \label{eqn:disk}
  \disk(Y, X) & = T \left( \lambda \begin{bmatrix} 0 & 0 \\ 0 & I_{n-k}
    \end{bmatrix} - \begin{bmatrix} I_{k} & 0 \\ 0 & 0 \end{bmatrix} \right)
    T^{-1},
\end{align}
with $T$, the right transformation matrix from~\eqref{eqn:weierform}.
The computation follows the inverse-free iteration~\cite{BaiDG97, Ben97a}
and a subspace extraction method~\cite{SunQ04, Ben11}.

In the following subsections, the ideas of the additive decomposition for two
general system classes are quickly summarized.


\subsection{Standard system case}

Assume a continuous-time standard system
\begin{align} \label{eqn:stdsys}
  \begin{aligned}
    \dot{x}(t) & = Ax(t) + Bu(t),\\
    y(t) & = Cx(t) + Du(t),
  \end{aligned}
\end{align}
with $A \in \mathbb{R}^{n \times n}$, $B \in \mathbb{R}^{n \times m}$,
$C \in \mathbb{R}^{p \times m}$, $D \in \mathbb{R}^{p \times m}$,
$A$ having no eigenvalues on the imaginary axis and its representation in the
frequency domain by the corresponding transfer function
\begin{align*}
  G(s) & = C(sI_{n} - A)^{-1}B + D,
\end{align*}
for $s \in \mathbb{C}$.
Most model reduction methods can only be applied to asymptotically stable
systems, which means in case of~\eqref{eqn:stdsys} that $A$ has only eigenvalues
with negative real parts.
Nevertheless, model reduction methods can be applied by decomposing the
system~\eqref{eqn:stdsys} into two subsystems, where the system matrices contain
either the stable or anti-stable system part, i.e., we search for a
transformation matrix $T$ such that
\begin{align*}
  T^{-1}AT & = \begin{bmatrix} A_{s} & 0 \\ 0 & A_{u} \end{bmatrix},
\end{align*}
where $A_{s}$ contains only the stable and $A_{u}$ the anti-stable eigenvalues.
Using $T$ as state-space transformation and partitioning accordingly the input
and output matrices yields the additive system decomposition of the system's
transfer function
\begin{align*}
  G(s) & = G_{s}(s) + G_{u}(s).
\end{align*}

Applying the matrix sign function~\eqref{eqn:sign} to $A$ gives the appropriate
spectral splitting, where the spectral projectors onto the deflating subspaces
are given as
\begin{align*}
  \begin{aligned}
    \cP_{s} & = \frac{1}{2}(I_{n} - \sign(A)) & \text{and} &&
      \cP_{u} & = \frac{1}{2}(I_{n} + \sign(A)).
  \end{aligned}
\end{align*}
Let $QR\Pi^{\trans} = I_{n} - \sign(A)$ be a pivoted QR decomposition, the 
dimension of the deflating subspace corresponding to the eigenvalues with
negative real part is given by $0.5(n + \trace(\sign(A)))$ and we get
\begin{align*}
  Q^{\trans}AQ & = \begin{bmatrix} A_{s} & W_{A} \\ 0 & A_{u} \end{bmatrix}.
\end{align*}
By solving the standard Sylvester equation
\begin{align} \label{eqn:sylv}
  -A_{u}X + XA_{s} - W_{A} & = 0,
\end{align}
the final transformation matrix and its inverse are given by
\begin{align} \label{eqn:adtftrafo}
  \begin{aligned}
    T & = Q \begin{bmatrix} I_{k} & X \\ 0 & I_{n-k} \end{bmatrix} & \text{and}
     && T^{-1} & = \begin{bmatrix} I_{k} & -X \\ 0 & I_{n-k} \end{bmatrix}
     Q^{\trans}.
  \end{aligned}
\end{align}

The \morlab{} implementation uses the Newton iteration with Frobenius norm
scaling for the computation of the matrix sign function as well as a matrix sign
function-based solver for the Sylvester equation~\eqref{eqn:sylv}.
Note that the actual transformation matrix~\eqref{eqn:adtftrafo} is never 
setup completely but only applied block wise on the original system to avoid
unnecessary computations.

\begin{remark}[Splitting of discrete-time standard systems]
  In case of discrete-time standard systems, the implementation involves the
  matrix sign function of $(A + I_{n})^{-1}(A - I_{n})$ and the solution of the
  discrete-time Sylvester equation $A_{u}^{-1}XA_{s} - X - A_{u}^{-1}W_{A}
  = 0$ for doing the spectral splitting with respect to the unit circle.
\end{remark}


\subsection{Descriptor system case}

Now, we consider the case of continuous-time descriptor systems
\begin{align} \label{eqn:descsys}
  \begin{aligned}
    E\dot{x}(t) & = Ax(t) + Bu(t),\\
    y(t) & = Cx(t) + Du(t),
  \end{aligned}
\end{align}
with $E, A \in \mathbb{R}^{n \times n}$, $B \in \mathbb{R}^{n \times m}$,
$C \in \mathbb{R}^{p \times m}$, $D \in \mathbb{R}^{p \times m}$,
$\lambda E - A$ having no finite eigenvalues on the imaginary axis and its
representation in the frequency domain by the corresponding transfer function
\begin{align*}
  \begin{aligned}
    G(s) & = C(sE - A)^{-1}B + D, && s \in \mathbb{C}.
  \end{aligned}
\end{align*}
In contrast to the previous section, an additional splitting for the algebraic
part corresponding to the infinite eigenvalues of $\lambda E - A$ is necessary,
i.e., we search for transformation matrices $W, T$ such that
\begin{align} \label{eqn:descsplit}
  W(\lambda E - A)T & = \lambda \begin{bmatrix} E_{s} & 0 & 0 \\ 0 & E_{u} & 0
    \\ 0 & 0 & E_{\infty} \end{bmatrix} - \begin{bmatrix} A_{s} & 0 & 0 \\
    0 & A_{u} & 0 \\ 0 & 0 & A_{\infty} \end{bmatrix},
\end{align}
where $\lambda E_{s} - A_{s}$ contains the finite stable eigenvalues,
$\lambda E_{u} - A_{u}$ the finite anti-stable eigenvalues and
$\lambda E_{\infty} - A_{\infty}$ only infinite eigenvalues.
Then, the system and its transfer function accordingly decouple into the
different parts
\begin{align*}
  G(s) & = G_{s}(s) + G_{u}(s) + G_{\infty}(s),
\end{align*}
as shown in Fig.~\ref{fig:morlab_workflow}.
For this purpose, the Theorem~3 from~\cite{morBenW18c} is used to construct
block wise orthogonal transformation matrices.

First, the splitting of the algebraic part is performed as
$G = G_{su} + G_{\infty}$ by using the matrix disk function.
In fact, the inverse-free iteration is applied to the matrix
pencil $\lambda (\alpha A) - E$ for appropriate scaling parameter $\alpha$
to compute matrices $\tA$ and $\tE$, whose null spaces are the
deflating subspaces of $\lambda (\alpha A) - E$ corresponding to the 
eigenvalues inside and outside the unit circle, respectively;
see~\cite{BaiDG97, Ben97a}.
Using a stabilized subspace extraction method~\cite{SunQ04, Ben11}, the
orthogonal projection matrices can be obtained and according
to~\cite{morBenW18c} combined into appropriate transformation matrices to get
\begin{align*}
  \tW (\lambda E - A) \tT & = \lambda \begin{bmatrix} E_{su} & 0 \\ 0 &
    E_{\infty} \end{bmatrix} - \begin{bmatrix} A_{su} & 0 \\ 0 &
    A_{\infty} \end{bmatrix},
\end{align*}
where $\lambda E_{su} - A_{su}$ contains all the finite eigenvalues.
Afterwards, the generalized matrix sign function, working implicitly on the
spectrum of $E_{su}^{-1}A_{su}$, is used such that the null spaces of
$E_{su} - \sign(A_{su}, E_{su})$ and $E_{su} + \sign(A_{su}, E_{su})$
are the deflating subspaces corresponding to the eigenvalues left and right
of the imaginary axis, respectively.
Using the same subspace extraction method and block transformation, the
block diagonalization~\eqref{eqn:descsplit} is accomplished.

\begin{remark}[Splitting of discrete-time descriptor systems]
  In the discrete-time descriptor case, the second splitting with respect
  to the imaginary axis needs to be replaced by a splitting with respect to the 
  unit disk.
  Although, this is the actual nature of the matrix disk function, for
  performance reasons, the generalized matrix sign function is used as
  $\sign(A_{su} - E_{su}, A_{su} + E_{su})$ replaces the sign functions above.
\end{remark}


\section{Model reduction with the MORLAB toolbox}%
\label{sec:mormeths}

Most of the model reduction methods in \morlab{} belong to the class of
projection-based model reduction, i.e., we are searching for truncation matrices
$W, T \in \mathbb{R}^{n \times r}$, which are used to project the state-space,
$x \approx T\hx$, and the corresponding equations.
For example, given a continuous-time descriptor system~\eqref{eqn:descsys},
the reduced-order system is computed by
\begin{align} \label{eqn:foproj}
  \begin{aligned}
    \underbrace{W^{\trans}ET}_{\hE}\dot{\hx}(t) & =
      \underbrace{W^{\trans}AT}_{\hA}\hx(t) +
      \underbrace{W^{\trans}B}_{\hB}u(t),\\
      \hy(t) & = \underbrace{CT}_{\hC}\hx(t) + \underbrace{D}_{\hD}u(t),
  \end{aligned}
\end{align}
with $\hE, \hA \in \mathbb{R}^{r \times r}$, $\hB \in \mathbb{R}^{r \times m}$,
$\hC \in \mathbb{R}^{p \times r}$ and $\hD = D$.
In the following, a very brief overview about the implemented model reduction
methods in \morlab{} is provided.


\subsection{First-order methods}

For the sake of generality in the \morlab{} setting, only the method
abbreviations are mentioned here.
According to the naming scheme, see Section~\ref{sec:design} and
Fig.~\ref{fig:morlab_func}, the abbreviations have to be connected with the 
system classes to give the actual \morlab{} function.

One of the oldest ideas for model reduction, and fitting with the spectral
splitting approach from before, is modal truncation.
While originally a part of the eigenvector basis was used for the
projection~\cite{morDav66}, the deflating subspaces from
Section~\ref{sec:specdecomp} are an appropriate choice when using
shifting and scaling on the spectrum of the system matrices.

A large part of the model reduction methods in \morlab{} are so-called
balancing-related methods.
In classical balanced truncation~\cite{morMoo81}, the continuous-time
Lyapunov equations
\begin{align} \label{eqn:gramians}
  \begin{aligned}
    AP + PA^{\trans} + BB^{\trans} & = 0,\\
    A^{\trans}Q + QA + C^{\trans}C & = 0,
  \end{aligned}
\end{align}
are solved for the system Gramians $P$ and $Q$, which are then used by, e.g.,
the square root or balancing-free square root method to compute the
reduced-order projection matrices; see, e.g.,~\cite{morVar91, morMehS05}.
The balancing-related methods are based on the idea of balanced truncation
but replace the Lyapunov equations~\eqref{eqn:gramians} by other matrix
equations, which infuse different properties to the resulting methods.
Some comments on the implementation of balancing-related methods in \morlab{}
are given for previous versions in~\cite{morBenW18a} for the standard system
case and the general idea of the implementation of model reduction for
descriptor systems is given in~\cite{morBenW18}.

\begin{table}[!t]
  \caption{First-order model reduction methods.}
  \label{tab:bt}
  \centering
  \setlength{\tabcolsep}{0pt}
  \begin{tabular}{lclc}
    \multicolumn{1}{c}{\textbf{Method}} &
      \multicolumn{1}{c}{\textbf{Routine name}} &
      \multicolumn{1}{c}{\textbf{Comment}} &
      \multicolumn{1}{c}{\textbf{References}}\\
      \hline\noalign{\medskip}
    Balanced truncation &
      \texttt{bt} &
      - preserves stability &
      \cite{morMoo81, morMehS05} \\
    Balanced stochastic truncation &
      \texttt{bst} &
      - preserves minimal phase &
      \cite{morGre88b, morBenS17}\\
    Frequency-limited balanced truncation &
      \texttt{flbt} &
      - local frequency approx. &
      \cite{morGawJ90, morImrG15}\\
    Time-limited balanced truncation &
      \texttt{tlbt} &
      - local time approx. &
      \cite{morGawJ90, morHaiGIetal17}\\
    LQG balanced truncation &
      \texttt{lqgbt} &
      - unstable system reduction &
      \cite{morJonS83, morBenS17}\\
    $\mathcal{H}_{\infty}$ balanced truncation &
      \texttt{hinfbt} &
      - unstable system reduction &
      \cite{morMusG91}\\
    Positive-real balanced truncation &
      \texttt{prbt} &
      - preserves passivity &
      \cite{morDesP84, morReiS10}\\
    Bounded-real balanced truncation &
      \texttt{brbt} &
      - preserves contractivity &
      \cite{morOpdJ88, morReiS10}\\ \noalign{\medskip}\hline\noalign{\medskip}
    Modal truncation & 
      \texttt{mt} &
      - preserves spectrum parts &
      \cite{morDav66, morBenQ05}\\
    Hankel-norm approximation &
      \texttt{hna} &
      - best approx. in Hankel-norm &
      \cite{morGlo84, morBenW18c}\\
    \noalign{\medskip}\hline\noalign{\smallskip}
  \end{tabular}
\end{table}

Also, the Hankel-norm approximation is implemented.
This method is non-projection-based, i.e., by construction, there are no
$W, T$ fulfilling~\eqref{eqn:foproj} and also $\hD = D$ does not hold anymore.
This method solves the optimal approximation problem in the Hankel semi-norm
and is also a good guess for the $\mathcal{H}_{\infty}$ approximation
problem~\cite{morGlo84, morBenW18c}.
It can be seen as a refinement of the balanced truncation method, since it is
also based on the solution of~\eqref{eqn:gramians}.

As an overview for the current \morlab{} version, Table~\ref{tab:bt} shows
all the implemented model reduction methods for first-order continuous-time
systems, with their routine abbreviation, a comment on their properties
and references for the standard and descriptor versions.

\begin{remark}[Discrete-time model reduction methods]
  Currently, only the methods \texttt{mt}, \texttt{bt} and \texttt{lqgbt} have
  discrete-time implementations for the standard and descriptor system case.
  Discrete-time equivalents of the continuous-time matrix equations
  are solved for those methods.
\end{remark}


\subsection{Second-order methods}

In case of systems with second-order time derivatives, the toolbox implements
different structure-preserving approaches.
Given the system structure from Table~\ref{tab:sys}, the reduced-order models
will also have the form
\begin{align} \label{eqn:sosys}
  \begin{aligned}
    \hM\ddot{\hx}(t) & = -\hK\hx(t) - \hE\dot{\hx}(t)
      + \hB_{u} u(t),\\
     \hy(t) & = \hC_{p}\hx(t) + \hC_{v}\dot{\hx}(t) + \hD u(t),
  \end{aligned}
\end{align}
with $\hM\, \hE, \hK \in \mathbb{R}^{r \times r}$,
$\hB_{u} \in \mathbb{R}^{r \times m}$, $\hC_{p},
\hC_{v} \in \mathbb{R}^{p \times r}$ and $\hD \in \mathbb{R}^{p \times m}$.
\morlab{} implements the second-order balanced truncation and balancing-related
methods for this purpose.
Originating in~\cite{morMeyS96, morReiS08, morChaLVetal06}, the second-order
balanced truncation approach uses a first-order realization of the original
second-order system and then restricts to parts of the system Gramians
to result in~\eqref{eqn:sosys}.
In~\cite{morBenW20b}, a collection of the different construction formulas
can be found that are all implemented in \morlab{}, as well as the frequency-
and time-limited second-order balanced truncation methods, which are also
implemented in \morlab{}.
The naming of the methods follows the previous subsection.


\section{Numerical examples}%
\label{sec:examples}

In the following, two benchmark examples are shown to demonstrate possible
applications of the \morlab{} toolbox.
The experiments reported here have been executed on  a machine with 2 Intel(R)
Xeon(R) Silver 4110 CPU processors running at 2.10GHz and equipped with
192 GB total main memory.
The computer is running on CentOS Linux release 7.5.1804 (Core) and using
MATLAB 9.4.0.813654 (R2018a) and the MORLAB toolbox version
5.0~\cite{morBenW19b}.

\begin{center}
  \fbox{%
    \parbox{.9\textwidth}{%
      The source code of the implementations used to compute the presented
      results can be obtained from
      \begin{center} 
        \url{https://doi.org/10.5281/zenodo.3678213}
      \end{center}
      and is authored by Jens Saak and Steffen W. R. Werner.
    }%
  }%
\end{center}


\subsection{Butterfly gyroscope}%
\label{sec:butterfly_gyro}

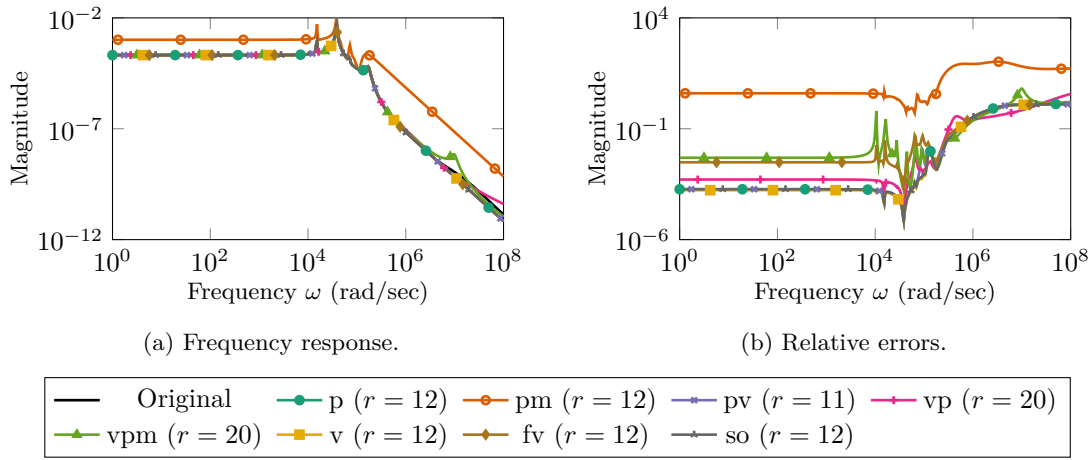
\begin{figure}[!t]
  \begin{center}
    \begin{subfigure}[t]{.49\textwidth}
      \begin{center}
  \tikzexternalenable%
  \tikzsetnextfilename{butterfly_gyro_tf}%
  \filemodCmp{graphics/butterfly_gyro_tf.tikz}{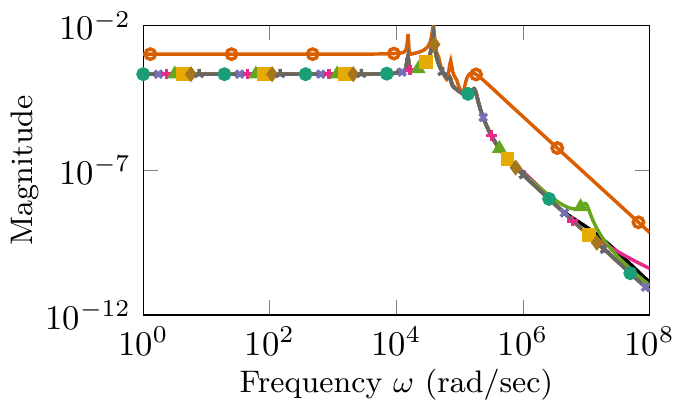}%
    {\tikzset{external/remake next}}{}%
  \begin{tikzpicture}  
  \pgfplotstableread{graphics/data/butterfly_gyro_tf.dat}\tableTF

  \begin{loglogaxis}[%
    width  = .7\textwidth,
    height = .4\textwidth,
    scale only axis,
    xmin = 1e+0,
    xmax = 1e+8,
    xtick = {1e+0, 1e+2, 1e+4, 1e+6, 1e+8},
    ymin = 1e-12,
    ymax = 1e-2,
    xminorticks = false,
    yminorticks = false,
    xlabel = {\small Frequency $\omega$ (rad/sec)},
    xlabel style = {yshift = .3em},
    ylabel = {\small Magnitude},
    ylabel style = {yshift = -.5em},
    scaled x ticks     = false,
    x tick label style = {/pgf/number format/fixed},
    cycle list name    = plotlist]
    
    \foreach \yIndex in {1, 2, ..., 9} {
      \addplot table[x index = 0, y index = \yIndex] {\tableTF};}
  \end{loglogaxis}
\end{tikzpicture}%
  \tikzexternaldisable%

        \subcaption{Frequency response.}
      \end{center}
    \end{subfigure}%
    \hfill%
    \begin{subfigure}[t]{.49\textwidth}
      \begin{center}
  \tikzexternalenable%
  \tikzsetnextfilename{butterfly_gyro_relerr}%
  \filemodCmp{graphics/butterfly_gyro_relerr.tikz}{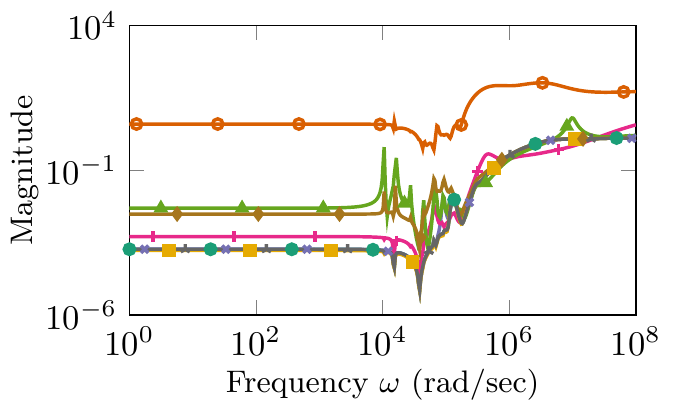}%
    {\tikzset{external/remake next}}{}%
  \begin{tikzpicture}  
  \pgfplotstableread{graphics/data/butterfly_gyro_relerr.dat}\tableERR

  \begin{loglogaxis}[%
    width  = .7\textwidth,
    height = .4\textwidth,
    scale only axis,
    xmin = 1e+0,
    xmax = 1e+8,
    xtick = {1e+0, 1e+2, 1e+4, 1e+6, 1e+8},
    ymin = 1e-6,
    ymax = 1e+4,
    xminorticks = false,
    yminorticks = false,
    xlabel = {\small Frequency $\omega$ (rad/sec)},
    xlabel style = {yshift = .3em},
    ylabel = {\small Magnitude},
    ylabel style = {yshift = -.5em},
    scaled x ticks     = false,
    x tick label style = {/pgf/number format/fixed},
    cycle list name    = plotlist]
    
    \pgfplotsset{cycle list shift = 1}
    
    \foreach \yIndex in {1, 2, ..., 8} {
      \addplot table[x index = 0, y index = \yIndex] {\tableERR};}
  \end{loglogaxis}
\end{tikzpicture}%
  \tikzexternaldisable%

        \subcaption{Relative errors.}
      \end{center}
    \end{subfigure}
    \vspace{.5\baselineskip}

  \tikzexternalenable%
  \tikzsetnextfilename{butterfly_gyro_legend}%
  \filemodCmp{graphics/butterfly_gyro_legend.tikz}{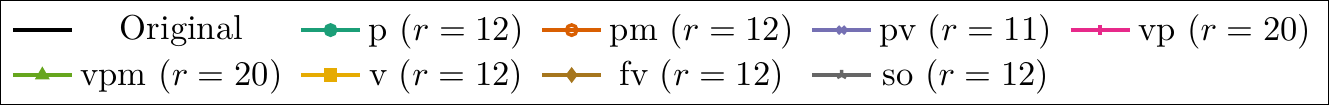}%
    {\tikzset{external/remake next}}{}%
  \begin{tikzpicture}
  \begin{axis}[%
    hide axis,
    scale only axis,
    width = 1mm,
    legend columns = 5,
    cycle list name = plotlist,
    legend style = {
      at     = {(0,0)},
      anchor = center,
      /tikz/every even column/.append style = {column sep = 0.1cm}}]
    
    \pgfplotsinvokeforeach{1,...,9}{\addplot coordinates {(0,0)};}
        
    \addlegendentry{Original};
    \addlegendentry{p ($r = 12$)};
    \addlegendentry{pm ($r = 12$)};
    \addlegendentry{pv ($r = 11$)};
    \addlegendentry{vp ($r = 20$)};
    \addlegendentry{vpm ($r = 20$)};
    \addlegendentry{v ($r = 12$)};
    \addlegendentry{fv ($r = 12$)};
    \addlegendentry{so ($r = 12$)};
  \end{axis}
\end{tikzpicture}%
  \tikzexternaldisable%

    \caption{Frequency domain results for the butterfly gyroscope.}
    \label{fig:butterfly_gyro}
  \end{center}
\end{figure}

As a first numerical example, we consider the butterfly gyroscope benchmark
example from~\cite{morwiki_gyro}; see~\cite{morBil05} for the background.
We will use the \morlab{} toolbox as backend software for a two-step model
reduction approach.
Thereby, a fast pre-reduction step is used to create an accurate, medium-scale
approximation of the original model and, afterwards, more sophisticated
model reduction methods are used to construct the final reduced-order model, 
see, e.g.,~\cite{morLehE07,  morSaaSW19}.
The model we consider now involves second-order time derivatives as it has
the form
\begin{align*}
  M\ddot{x}(t) + E\dot{x}(t) + Kx(t) & = B_{u}u(t),\\
  y(t) & = C_{p}x(t),
\end{align*}
with a state-space dimension $n = 17\,361$ and $m = 1$, $p = 12$ inputs and
outputs, respectively.

As in~\cite{morSaaSW19}, we use the structure-preserving interpolation
framework from~\cite{morBeaG09} as efficient pre-reduction method that preserves
the system structure in the intermediate medium-scale approximation.
We compute a single projection basis the same way as in~\cite{morSaaSW19}
using the sampling points $\pm\texttt{logspace(0, 8, 100)}i$.
After orthogonalization by the economy size QR decomposition, the intermediate
reduced-order model has the state-space dimension $2\,600$.
Now, we apply the second-order balanced truncation methods from \morlab{} to the 
intermediate model.
The toolbox supports an all-at-once approach for balancing-related model
reduction, i.e., the underlying Gramians are computed once and then used for
several different reduced-order models.
Therefore, we can compute all $8$ different second-order balancing formulas
from~\cite{morBenW20b} at the same time and compare them afterwards.

Fig.~\ref{fig:butterfly_gyro} shows the resulting reduced-order models
in the frequency domain.
The relative error was computed
\begin{align*}
  \frac{\lVert G(i\omega) - \hG(i\omega)\rVert_{2}}%
    {\lVert G(i\omega)\rVert_{2}},
\end{align*}
in the frequency range $\omega \in [10^{0}, 10^{8}]$.
Both plots were directly generated with the \morlab{} routine
\texttt{ml\_sigmaplot}, which computes sigma and error plots for an arbitrary
number of given models.
The notation in the legend follows the formulas from~\cite{morBenW20b}.
Except for the pm and vpm models, all other reduced-order models are stable.
Clearly, the winners are p, v, pv and so, which all have basically the same size
and error behavior.


\subsection{Parametric thermal block model}%
\label{sec:thermalblock}

As second example, we consider the parametric thermal block model as described
in~\cite{morwiki_thermalblock} with the single parameter setup.
Following this description, we consider the first-order generalized state-space
system
\begin{align} \label{eqn:thermalblock}
  \begin{aligned}
    E\dot{x}(t;\mu) & = A(\mu)x(t;\mu) + Bu(t),\\
    y(t;\mu) & = Cx(t;\mu),
  \end{aligned}
\end{align}
where $A(\mu) = A_{0} + \mu \left(0.2 A_{1} + 0.4 A_{2} + 0.6 A_{3} +
0.8 A_{4} \right),$ with the parameter $\mu \in [10^{-6}, 10^{2}]$, the 
state-space dimension $n = 7\,488$ and $m = 1$, $p = 4$ inputs and outputs,
respectively.
The matrix pencil $\lambda E - A(\mu)$ is finite and stable for all parameter
values $\mu$ in the range of interest.

Although \morlab{} does not implement parametric system classes yet, we want to
use the toolbox as model reduction backend for two-step parametric
model reduction methods.
The first idea is taken from~\cite{morBauB09}.
Given some non-parametric reduced-order models $G_{j}$ computed for parameter
samples $\mu_{j}$, $j = 1, \ldots, k$, a global parameter interpolating
system can be constructed in the frequency domain using Lagrange interpolation
as
\begin{align} \label{eqn:interptf}
  \hG(s, \mu) = \sum_{j = 1}^{k} \ell_{j}(\mu)G_{j}(s),
\end{align}
with $\ell_{j}(\mu)$ Lagrange basis functions in the parameter $\mu$ with the
knot vector $\mu_{1}, \ldots, \mu_{k}$.
Rewriting the sum~\eqref{eqn:interptf} gives a realization for the
interpolating reduced-order model
\begin{align*}
  \begin{aligned}
    \hE & = \begin{bmatrix} \hE_{1} & & \\ & \mddots & \\ & & \hE_{k}
      \end{bmatrix}, &
    \hA & = \begin{bmatrix} \hA_{1} & & \\ & \mddots & \\ & & \hA_{k}
      \end{bmatrix},\\
    \hB & = \begin{bmatrix} \hB_{1} \\ \mvdots \\ \hB_{k} \end{bmatrix}, &
    \hC & = \begin{bmatrix} \ell_{1}(\mu)\hC_{1}, & \hdots, &
      \ell_{k}(\mu)\hC_{k} \end{bmatrix},
  \end{aligned}
\end{align*}
where $\hE_{j}, \hA_{j}, \hB_{j}, \hC_{j}$ are the matrices of the local
reduced-order models.
Thinking of other scalar function approximation methods, easy extensions
of~\eqref{eqn:interptf} come into mind.
Replacing the Lagrange basis functions $\ell_{j}(\mu)$ by linear B-splines
$b_{1, j}(\mu)$ over the knot vector $\mu_{1}, \ldots, \mu_{k}$, we can
construct a piecewise linear interpolating reduced-order model.
Another idea would be to use the variation diminishing B-spline approximation,
which just needs some modifications of the knot vector used for the basis
functions.
In general, this transfer function interpolation-based approach comes with
several advantages.
First, it does not matter how the local reduced-order models were computed or
which size they have.
If all local reduced-order models were stable, the global interpolating
one will be stable by construction, too.
Also, instead of setting up the complete reduced-order model, it can be
advantageous to use the local reduced-order models for simulations in parallel
and combine the results at the end by the parametric output matrix.

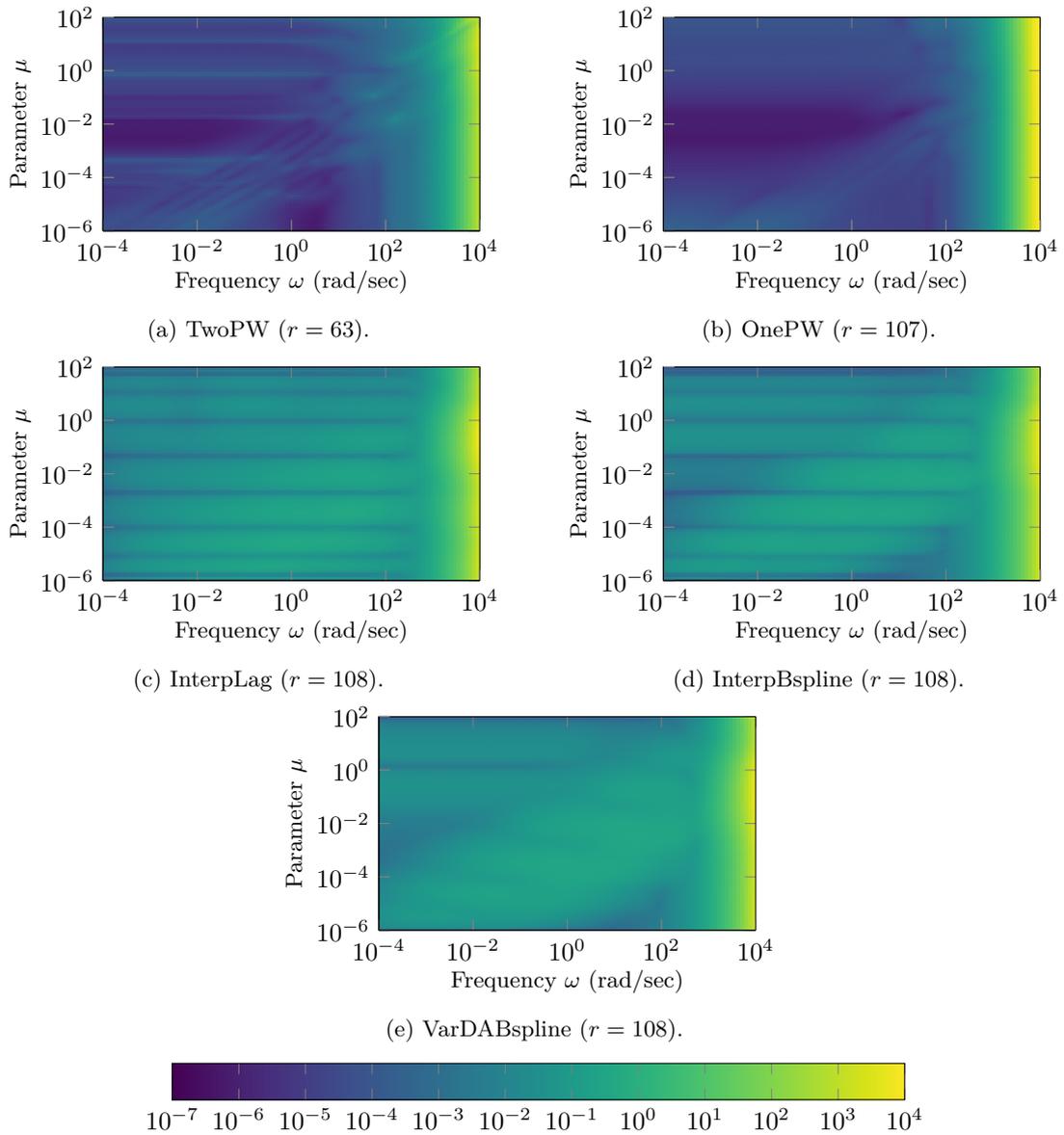
\begin{figure}[!t]
  \begin{center}
    \begin{subfigure}[t]{.49\textwidth}
      \begin{center}
  \tikzexternalenable%
  \tikzsetnextfilename{piecewise_ts_bt_err_rel}%
  \filemodCmp{graphics/piecewise_ts_bt_err_rel.tikz}{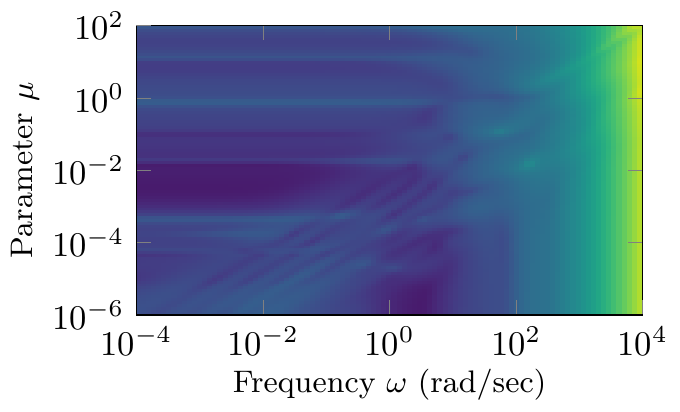}%
    {\tikzset{external/remake next}}{}%
  \begin{tikzpicture}
  \pgfplotstableread{graphics/data/piecewise_ts_bt_all.dat}\tableROM
  
  \begin{loglogaxis}[
    view={0}{90},
    width  = .7\textwidth,
    height = .4\textwidth,
    scale only axis,
    xmin = 1e-4,
    xmax = 1e+4,
    ymin = 1e-6,
    ymax = 1e+2,
    xtick = {1e-4, 1e-2, 1e0, 1e+2, 1e+4},
    ytick = {1e-6, 1e-4, 1e-2, 1e+0, 1e+2},
    zmode = log,
    log base z     = 10,
    point meta min = -7,
    point meta max = 4,
    mesh/ordering = y varies,
    mesh/rows = 100,
    mesh/cols = 100,
    xlabel = {\small Frequency $\omega$ (rad/sec)},
    xlabel style = {yshift = .3em},
    ylabel = {\small Parameter $\mu$},
    ylabel style = {yshift = -.3em},
    scaled x ticks = false,
    x tick label style = {/pgf/number format/fixed}]
        
      \addplot3[surf, shader = flat]
        table[x index = 0, y index = 1, z index = 4] {\tableROM};
            
  \end{loglogaxis}
\end{tikzpicture}%
  \tikzexternaldisable%

        \subcaption{TwoPW ($r = 63$).}
      \end{center}
    \end{subfigure}%
    \hfill%
    \begin{subfigure}[t]{.49\textwidth}
      \begin{center}
  \tikzexternalenable%
  \tikzsetnextfilename{piecewise_os_bt_err_rel}%
  \filemodCmp{graphics/piecewise_os_bt_err_rel.tikz}{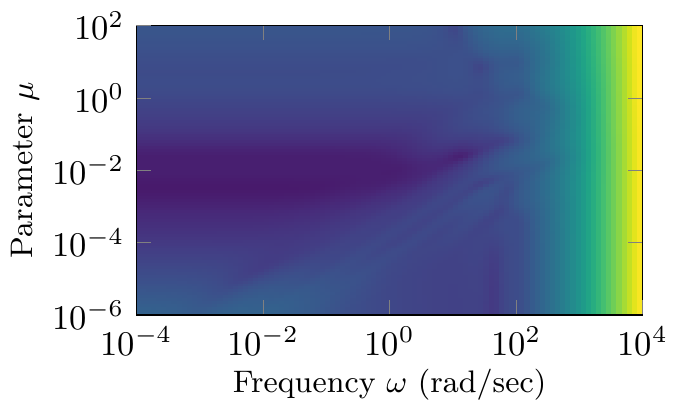}%
    {\tikzset{external/remake next}}{}%
  \begin{tikzpicture}
  \pgfplotstableread{graphics/data/piecewise_os_bt_all.dat}\tableROM
  
  \begin{loglogaxis}[
    view={0}{90},
    width  = .7\textwidth,
    height = .4\textwidth,
    scale only axis,
    xmin = 1e-4,
    xmax = 1e+4,
    ymin = 1e-6,
    ymax = 1e+2,
    xtick = {1e-4, 1e-2, 1e0, 1e+2, 1e+4},
    ytick = {1e-6, 1e-4, 1e-2, 1e+0, 1e+2},
    zmode = log,
    log base z     = 10,
    point meta min = -7,
    point meta max = 4,
    mesh/ordering = y varies,
    mesh/rows = 100,
    mesh/cols = 100,
    xlabel = {\small Frequency $\omega$ (rad/sec)},
    xlabel style = {yshift = .3em},
    ylabel = {\small Parameter $\mu$},
    ylabel style = {yshift = -.3em},
    scaled x ticks = false,
    x tick label style = {/pgf/number format/fixed}]
        
      \addplot3[surf, shader = flat]
        table[x index = 0, y index = 1, z index = 4] {\tableROM};
            
  \end{loglogaxis}
\end{tikzpicture}%
  \tikzexternaldisable%

        \subcaption{OnePW ($r = 107$).}
      \end{center}
    \end{subfigure}
    
    \begin{subfigure}[t]{.49\textwidth}
      \begin{center}
  \tikzexternalenable%
  \tikzsetnextfilename{param_interp_bt_err_rel}%
  \filemodCmp{graphics/param_interp_bt_err_rel.tikz}{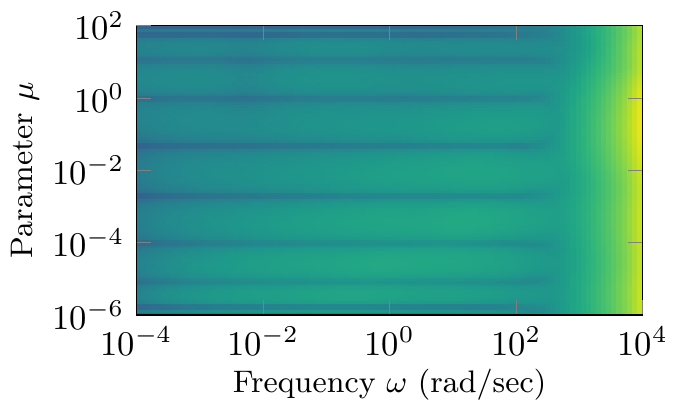}%
    {\tikzset{external/remake next}}{}%
  \begin{tikzpicture}
  \pgfplotstableread{graphics/data/param_interp_bt_all.dat}\tableROM
  
  \begin{loglogaxis}[
    view={0}{90},
    width  = .7\textwidth,
    height = .4\textwidth,
    scale only axis,
    xmin = 1e-4,
    xmax = 1e+4,
    ymin = 1e-6,
    ymax = 1e+2,
    xtick = {1e-4, 1e-2, 1e0, 1e+2, 1e+4},
    ytick = {1e-6, 1e-4, 1e-2, 1e+0, 1e+2},
    zmode = log,
    log base z     = 10,
    point meta min = -7,
    point meta max = 4,
    mesh/ordering = y varies,
    mesh/rows = 100,
    mesh/cols = 100,
    xlabel = {\small Frequency $\omega$ (rad/sec)},
    xlabel style = {yshift = .3em},
    ylabel = {\small Parameter $\mu$},
    ylabel style = {yshift = -.3em},
    scaled x ticks = false,
    x tick label style = {/pgf/number format/fixed}]
        
      \addplot3[surf, shader = flat]
        table[x index = 0, y index = 1, z index = 4] {\tableROM};
            
  \end{loglogaxis}
\end{tikzpicture}%
  \tikzexternaldisable%

        \subcaption{InterpLag ($r = 108$).}
      \end{center}
    \end{subfigure}%
    \hfill%
    \begin{subfigure}[t]{.49\textwidth}
      \begin{center}
  \tikzexternalenable%
  \tikzsetnextfilename{param_bspline_lin_bt_err_rel}%
  \filemodCmp{graphics/param_bspline_lin_bt_err_rel.tikz}{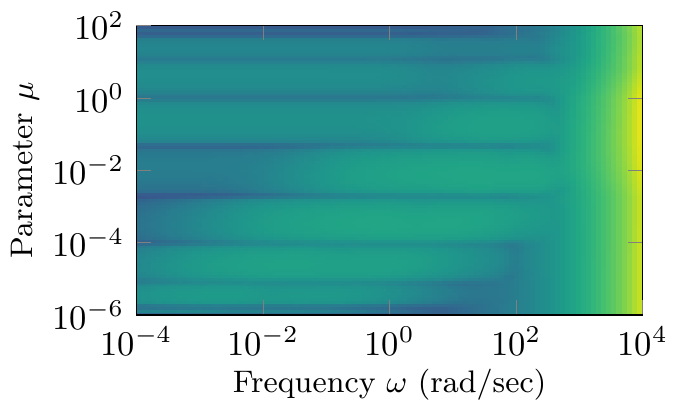}%
    {\tikzset{external/remake next}}{}%
  \begin{tikzpicture}
  \pgfplotstableread{graphics/data/param_bspline_lin_bt_all.dat}\tableROM
  
  \begin{loglogaxis}[
    view={0}{90},
    width  = .7\textwidth,
    height = .4\textwidth,
    scale only axis,
    xmin = 1e-4,
    xmax = 1e+4,
    ymin = 1e-6,
    ymax = 1e+2,
    xtick = {1e-4, 1e-2, 1e0, 1e+2, 1e+4},
    ytick = {1e-6, 1e-4, 1e-2, 1e+0, 1e+2},
    zmode = log,
    log base z     = 10,
    point meta min = -7,
    point meta max = 4,
    mesh/ordering = y varies,
    mesh/rows = 100,
    mesh/cols = 100,
    xlabel = {\small Frequency $\omega$ (rad/sec)},
    xlabel style = {yshift = .3em},
    ylabel = {\small Parameter $\mu$},
    ylabel style = {yshift = -.3em},
    scaled x ticks = false,
    x tick label style = {/pgf/number format/fixed}]
        
      \addplot3[surf, shader = flat]
        table[x index = 0, y index = 1, z index = 4] {\tableROM};
            
  \end{loglogaxis}
\end{tikzpicture}%
  \tikzexternaldisable%

        \subcaption{InterpBspline ($r = 108$).}
      \end{center}
    \end{subfigure}
    
    \begin{subfigure}[t]{.49\textwidth}
      \begin{center}
  \tikzexternalenable%
  \tikzsetnextfilename{param_bspline_quad_bt_err_rel}%
  \filemodCmp{graphics/param_bspline_quad_bt_err_rel.tikz}{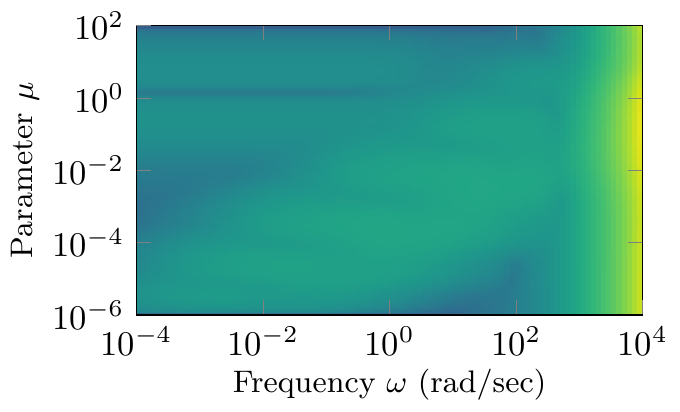}%
    {\tikzset{external/remake next}}{}%
  \begin{tikzpicture}
  \pgfplotstableread{graphics/data/param_bspline_quad_bt_all.dat}\tableROM
  
  \begin{loglogaxis}[
    view={0}{90},
    width  = .7\textwidth,
    height = .4\textwidth,
    scale only axis,
    xmin = 1e-4,
    xmax = 1e+4,
    ymin = 1e-6,
    ymax = 1e+2,
    xtick = {1e-4, 1e-2, 1e0, 1e+2, 1e+4},
    ytick = {1e-6, 1e-4, 1e-2, 1e+0, 1e+2},
    zmode = log,
    log base z     = 10,
    point meta min = -7,
    point meta max = 4,
    mesh/ordering = y varies,
    mesh/rows = 100,
    mesh/cols = 100,
    xlabel = {\small Frequency $\omega$ (rad/sec)},
    xlabel style = {yshift = .3em},
    ylabel = {\small Parameter $\mu$},
    ylabel style = {yshift = -.3em},
    scaled x ticks = false,
    x tick label style = {/pgf/number format/fixed}]
        
      \addplot3[surf, shader = flat]
        table[x index = 0, y index = 1, z index = 4] {\tableROM};
            
  \end{loglogaxis}
\end{tikzpicture}%
  \tikzexternaldisable%

        \subcaption{VarDABspline ($r = 108$).}
      \end{center}
    \end{subfigure}
    \vspace{.5\baselineskip}

  \tikzexternalenable%
  \tikzsetnextfilename{parametric_bt_legend}%
  \filemodCmp{graphics/parametric_bt_legend.tikz}{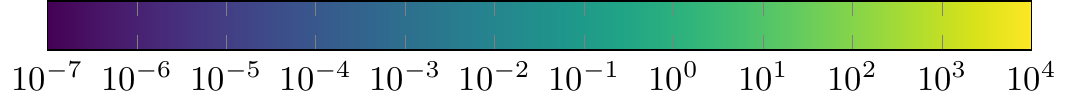}%
    {\tikzset{external/remake next}}{}%
  \begin{tikzpicture}
  \begin{axis}[%
    hide axis,
    scale only axis,
    width  = 10cm,
    height = .1cm,
    point meta min = -7,
    point meta max = 4,
    colorbar,
    colorbar horizontal,
    colorbar style = {
      xticklabel = $10^{\pgfmathparse{\tick}
        \pgfmathprintnumber\pgfmathresult}$,
      at = {(.5, 0)},
      anchor = north},
    scaled x ticks     = false,
    x tick label style = {/pgf/number format/fixed}]
  \end{axis}
\end{tikzpicture}%
  \tikzexternaldisable%

    \caption{Relative errors in the frequency domain of different parametric
      extensions for the thermal block model.}
    \label{fig:thermalblock_freq}
  \end{center}
\end{figure}

A different approach is given by the piecewise approximation; see,
e.g.,~\cite{morBauBBetal11}.
For this method, let the local reduced-order models be computed by projection
methods and the projection matrices be collected as
$W = [W_{1}, \ldots, W_{k}]$ and $T = [T_{1}, \ldots, T_{k}]$.
The parametric reduced-order system is then computed using $W, T$ as projection
matrices on the original system, as in~\eqref{eqn:foproj}.
Concerning the parametric matrix $A(\mu)$ in~\eqref{eqn:thermalblock},
we note that
\begin{align*}
  W^{\trans}A(\mu)T & = W^{\trans}A_{0}T + \mu \left(0.2 W^{\trans}A_{1}T +
    0.4 W^{\trans}A_{2}T + 0.6 W^{\trans}A_{3}T + 0.8 W^{\trans}A_{4}T \right)\\
  & = \hA_{0} + \mu \left(0.2 \hA_{1} + 0.4 \hA_{2} + 0.6 \hA_{3} + 0.8 \hA_{4}
    \right).
\end{align*}
Using this method, we can preserve the exact parameter dependency in the
reduced-order model.
Variants of it, for example, use column compression of $T$ and $W$ to control
the size of the resulting reduced-order model.
Also, it needs to be noted that by concatenation of the projection matrices,
original properties like stability preservation can be lost.
Therefore, modifications like a one-sided projection by combining $[W, T]$ into
a single basis can be used to handle most systems.

\begin{figure}[!t]
  \begin{center}
    \begin{subfigure}[t]{.49\textwidth}
      \begin{center}
  \tikzexternalenable%
  \tikzsetnextfilename{piecewise_os_bt_sim_err_rel}%
  \filemodCmp{graphics/piecewise_os_bt_sim_err_rel.tikz}{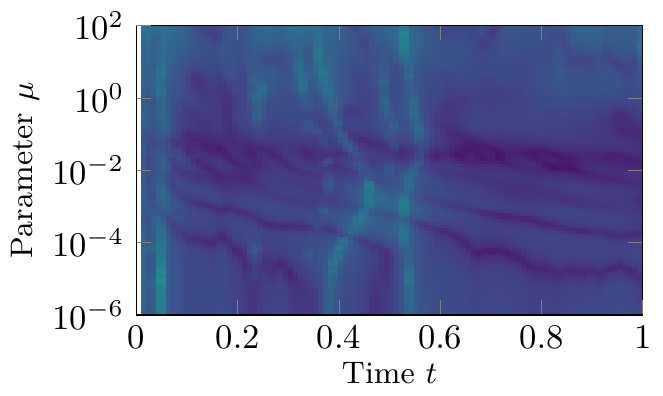}%
    {\tikzset{external/remake next}}{}%
  \begin{tikzpicture}
  \pgfplotstableread{graphics/data/piecewise_os_bt_sim_all.dat}\tableROM
  
  \begin{semilogyaxis}[
    view   = {0}{90},
    width  = .7\textwidth,
    height = .4\textwidth,
    scale only axis,
    xmin = 0,
    xmax = 1,
    ymin = 1e-6,
    ymax = 1e+2,
    ytick = {1e-6, 1e-4, 1e-2, 1e+0, 1e+2},
    zmode = log,
    log base z     = 10,
    point meta min = -6.3,
    point meta max = 3.9,
    mesh/ordering = y varies,
    mesh/rows = 101,
    mesh/cols = 100,
    xlabel = {\small Time $t$},
    xlabel style = {yshift = .3em},
    ylabel = {\small Parameter $\mu$},
    ylabel style = {yshift = -.3em},
    scaled x ticks = false,
    x tick label style = {/pgf/number format/fixed}]
        
      \addplot3[surf, shader = flat]
        table[x index = 0, y index = 1, z index = 3] {\tableROM};
            
  \end{semilogyaxis}
\end{tikzpicture}%
  \tikzexternaldisable%

        \subcaption{OnePW ($r = 107$).}
      \end{center}
    \end{subfigure}%
    \hfill%
    \begin{subfigure}[t]{.49\textwidth}
      \begin{center}
  \tikzexternalenable%
  \tikzsetnextfilename{param_interp_bt_sim_err_rel}%
  \filemodCmp{graphics/param_interp_bt_sim_err_rel.tikz}{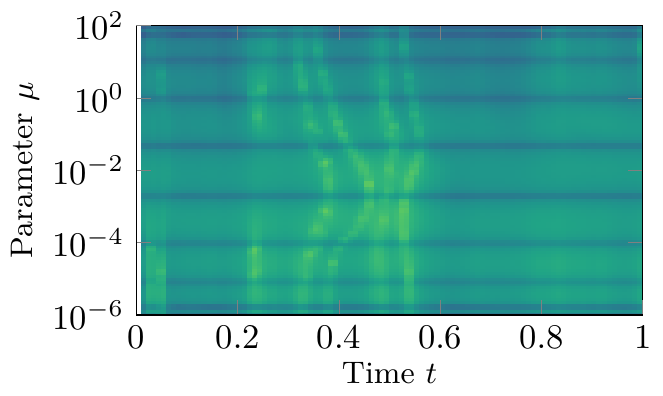}%
    {\tikzset{external/remake next}}{}%
  \begin{tikzpicture}
  \pgfplotstableread{graphics/data/param_interp_bt_sim_all.dat}\tableROM
  
  \begin{semilogyaxis}[
    view   = {0}{90},
    width  = .7\textwidth,
    height = .4\textwidth,
    scale only axis,
    xmin = 0,
    xmax = 1,
    ymin = 1e-6,
    ymax = 1e+2,
    ytick = {1e-6, 1e-4, 1e-2, 1e+0, 1e+2},
    zmode = log,
    log base z     = 10,
    point meta min = -6.3,
    point meta max = 3.9,
    mesh/ordering = y varies,
    mesh/rows = 101,
    mesh/cols = 100,
    xlabel = {\small Time $t$},
    xlabel style = {yshift = .3em},
    ylabel = {\small Parameter $\mu$},
    ylabel style = {yshift = -.3em},
    scaled x ticks = false,
    x tick label style = {/pgf/number format/fixed}]
        
      \addplot3[surf, shader = flat]
        table[x index = 0, y index = 1, z index = 3] {\tableROM};
            
  \end{semilogyaxis}
\end{tikzpicture}%
  \tikzexternaldisable%

        \subcaption{InterpLag ($r = 108$).}
      \end{center}
    \end{subfigure}
    
    \begin{subfigure}[t]{.49\textwidth}
      \begin{center}
  \tikzexternalenable%
  \tikzsetnextfilename{param_bspline_lin_bt_sim_err_rel}%
  \filemodCmp{graphics/param_bspline_lin_bt_sim_err_rel.tikz}{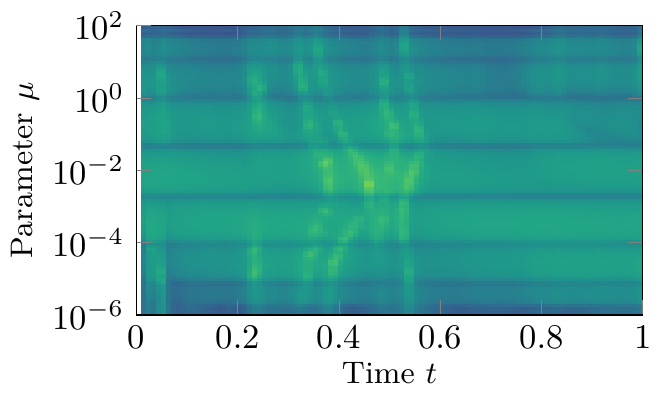}%
    {\tikzset{external/remake next}}{}%
  \begin{tikzpicture}
  \pgfplotstableread{graphics/data/param_bspline_lin_bt_sim_all.dat}\tableROM
  
  \begin{semilogyaxis}[
    view   = {0}{90},
    width  = .7\textwidth,
    height = .4\textwidth,
    scale only axis,
    xmin = 0,
    xmax = 1,
    ymin = 1e-6,
    ymax = 1e+2,
    ytick = {1e-6, 1e-4, 1e-2, 1e+0, 1e+2},
    zmode = log,
    log base z     = 10,
    point meta min = -6.3,
    point meta max = 3.9,
    mesh/ordering = y varies,
    mesh/rows = 101,
    mesh/cols = 100,
    xlabel = {\small Time $t$},
    xlabel style = {yshift = .3em},
    ylabel = {\small Parameter $\mu$},
    ylabel style = {yshift = -.3em},
    scaled x ticks = false,
    x tick label style = {/pgf/number format/fixed}]
        
      \addplot3[surf, shader = flat]
        table[x index = 0, y index = 1, z index = 3] {\tableROM};
            
  \end{semilogyaxis}
\end{tikzpicture}%
  \tikzexternaldisable%

        \subcaption{InterpBspline ($r = 108$).}
      \end{center}
    \end{subfigure}%
    \hfill%
    \begin{subfigure}[t]{.49\textwidth}
      \begin{center}
  \tikzexternalenable%
  \tikzsetnextfilename{param_bspline_quad_bt_sim_err_rel}%
  \filemodCmp{graphics/param_bspline_quad_bt_sim_err_rel.tikz}{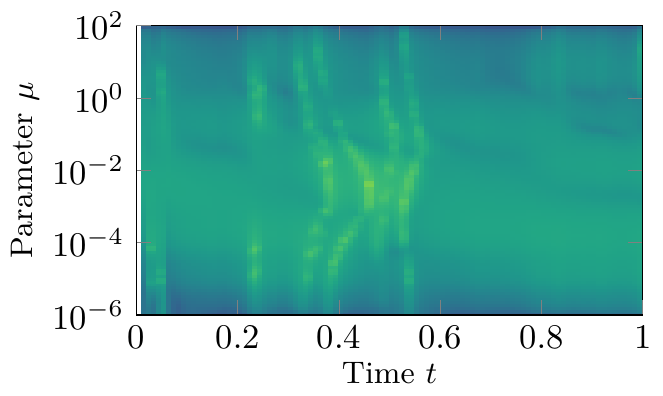}%
    {\tikzset{external/remake next}}{}%
  \begin{tikzpicture}
  \pgfplotstableread{graphics/data/param_bspline_quad_bt_sim_all.dat}\tableROM
  
  \begin{semilogyaxis}[
    view   = {0}{90},
    width  = .7\textwidth,
    height = .4\textwidth,
    scale only axis,
    xmin = 0,
    xmax = 1,
    ymin = 1e-6,
    ymax = 1e+2,
    ytick = {1e-6, 1e-4, 1e-2, 1e+0, 1e+2},
    zmode = log,
    log base z     = 10,
    point meta min = -6.3,
    point meta max = 3.9,
    mesh/ordering = y varies,
    mesh/rows = 101,
    mesh/cols = 100,
    xlabel = {\small Time $t$},
    xlabel style = {yshift = .3em},
    ylabel = {\small Parameter $\mu$},
    ylabel style = {yshift = -.3em},
    scaled x ticks = false,
    x tick label style = {/pgf/number format/fixed}]
        
      \addplot3[surf, shader = flat]
        table[x index = 0, y index = 1, z index = 3] {\tableROM};
            
  \end{semilogyaxis}
\end{tikzpicture}%
  \tikzexternaldisable%

        \subcaption{VarDABspline ($r = 108$).}
      \end{center}
    \end{subfigure}
    
    \vspace{.5\baselineskip}

  \tikzexternalenable%
  \tikzsetnextfilename{parametric_bt_sim_legend}%
  \filemodCmp{graphics/parametric_bt_sim_legend.tikz}{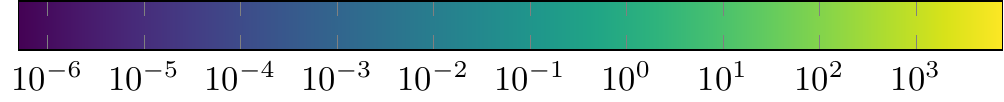}%
    {\tikzset{external/remake next}}{}%
  \begin{tikzpicture}
  \begin{axis}[%
    hide axis,
    scale only axis,
    width  = 10cm,
    height = .1cm,
    point meta min = -6.3,
    point meta max = 3.9,
    colorbar,
    colorbar horizontal,
    colorbar style = {
      xticklabel = $10^{\pgfmathparse{\tick}
        \pgfmathprintnumber\pgfmathresult}$,
      at = {(.5, 0)},
      anchor = north},
    scaled x ticks     = false,
    x tick label style = {/pgf/number format/fixed}]
  \end{axis}
\end{tikzpicture}%
  \tikzexternaldisable%

    \caption{Relative errors in the time simulation of different parametric
      extensions for the thermal block model.}
    \label{fig:thermalblock_sim}
  \end{center}
\end{figure}
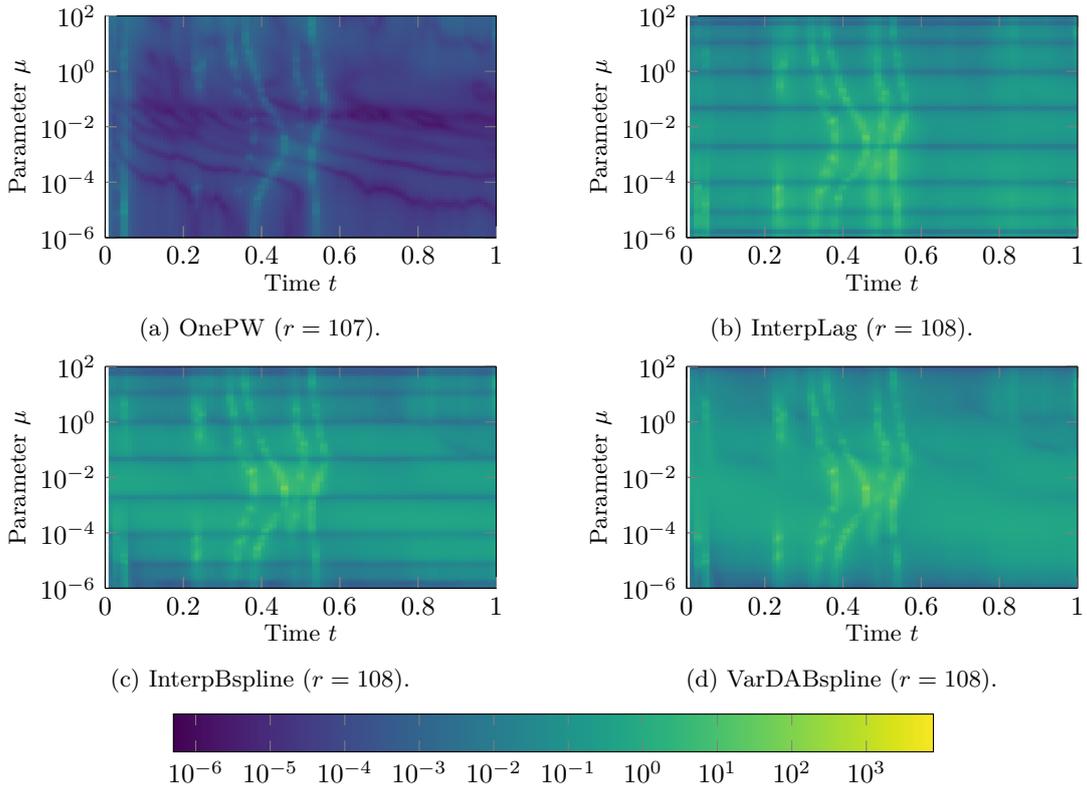

For our numerical example, we will use the following setup.
For the parameter sampling points, we use $10$ logarithmically distributed
Chebyshev roots, i.e., let $\nu_{1}, \ldots, \nu_{k}$ be the Chebyshev roots
in the interval $[-6, 2]$, the sampling points are given as
$\mu_{j} = 10^{\nu_{j}}$.
The local reduced-order models are computed by the balanced truncation routine
from \morlab{} (\texttt{ml\_ct\_dss\_bt}) using $10^{-4}$ for the absolute error
bound and we save the reduced-order models as well as the projection matrices
for the parametric approaches.
The following different parametric reduced-order models are then computed:
\begin{itemize}
  \item two-sided piecewise approximation (TwoPW), where the final truncated
    projection matrices were compressed using singular value decompositions and
    a relative truncation tolerance of $10^{-4}$,
  \item one-sided piecewise approximation (OnePW), where the final truncated
    projection matrix was compressed using the basis concatenation and the 
    singular value decomposition with relative truncation tolerance $10^{-4}$,
  \item transfer function interpolation using Lagrange basis functions
    (InterpLag),
  \item transfer function interpolation using linear B-splines (InterpBspline),
  \item transfer function approximation using the variation diminishing
    approximation with qua\-dra\-tic B-spline basis functions (VarDABspline).
\end{itemize}

Fig.~\ref{fig:thermalblock_freq} shows the results in the frequency domain,
where we computed the point wise relative errors as
\begin{align*}
  \frac{\lVert G(i\omega, \mu) - \hG(i\omega, \mu)\rVert_{2}}%
    {\lVert G(i\omega, \mu)\rVert_{2}},
\end{align*}
in the ranges $\omega \in [10^{-4}, 10^{4}]$ and $\mu \in [10^{-6}, 10^{2}]$.
The piecewise methods, TwoPW and OnePW, are the clear winners of the comparison.
We note that TwoPW is unstable for all parameters, while OnePW is stable.
Also, the interpolation approaches work nicely, where the interpolation property
is clearly visible in the plots.
The variation diminishing B-spline result, VarDABspline, seems to be a smoother
version of InterpBspline.

In the time domain, we simulate the parametric systems with using a pre-sampled
white noise input signal.
The relative errors shown in Fig.~\ref{fig:thermalblock_sim} are computed
by
\begin{align*}
  \sqrt{\sum\limits_{j = 1}^{4} \frac{\lvert y_{j}(t;\mu) - \hy_{j}(t;\mu)
    \rvert^{2}}{\lvert y_{j}(t;\mu) \rvert^{2}}}
\end{align*}
in the ranges $t \in [0, 1]$ and $\mu \in [10^{-6}, 10^{2}]$.
The TwoPW is not shown in Fig.~\ref{fig:thermalblock_sim}, since due to the
instability in all parameters, no useful results were computed during the
simulation.
For the rest, we see that again OnePW performs overall very good.
Also we see that the B-spline approaches and classical Lagrange interpolation
give more or less the same results.


\section{Conclusions}%
\label{sec:conclusions}

We presented the \morlab{} toolbox as efficient software solution for model
reduction of dense, medium-scale linear time-invariant systems.
We gave an overview about the main features and structure of the toolbox, as
well as underlying programming principles.
An important point when considering unstructured systems is the spectral
splitting, which we showed in \morlab{} to be based on spectral projection
methods.
Following the computational steps led to an overview about the implemented
model reduction methods in \morlab{}.
We gave two numerical examples to illustrate how \morlab{} can be used as
backend software for different system types.
In the first example, \morlab{} provided the efficient, structure-preserving
implementation of sophisticated model reduction methods that are used in
two-step approaches.
In the second example, we used \morlab{} to generate local reduced-order models
that were afterwards combined by different techniques to construct parametric
reduced-order systems.


\section*{Acknowledgment}%
\addcontentsline{toc}{section}{Acknowledgment}

This work was supported by the German Research Foundation (DFG) Research
Training Group 2297 ``MathCoRe'', Magdeburg.


\addcontentsline{toc}{section}{References}
\bibliographystyle{plainurl}
\bibliography{bibtex/myref}

\end{document}